\documentclass[pdflatex,notitlepage,showkeys,floatfix,aps,pra]{revtex4-2}
\usepackage{amsmath,amssymb,amsthm,physics}
\usepackage{bm}
\usepackage{graphicx}
\usepackage{hyperref}
\usepackage{geometry}
\usepackage{xfrac}
\usepackage{dsfont}
\usepackage{mathtools}
\usepackage[dvipsnames]{xcolor}

\begin{document}

\title{Qudit Implementation of the Rodeo Algorithm for Quantum Spectral Filtering}

\author{J. C. S. Rocha}
  \email{jcsrocha@ufop.edu.br}
  \affiliation{Departamento de Física, Universidade Federal de Ouro Preto - UFOP, Ouro Preto, Minas Gerais, Brasil.}
  
\author{R. A. Dias}
  \email{rodrigo.dias@ufjf.br}
  \affiliation{Departamento de Física, Universidade Federal de Juiz de Fora - UFJF, Minas Gerais, Brasil.}
  
\date{\today}

\begin{abstract}
Qudits, the multi-level generalization of qubits, provide a natural extension of the binary paradigm in quantum computation and offer new opportunities to enhance algorithmic performance. Beyond their direct applicability to the simulation of multi-level quantum systems, higher-dimensional ancillae can improve sampling efficiency in quantum algorithms by enabling the simultaneous implementation of multiple control operations, thereby reducing circuit complexity. In this work, we pursue three main objectives. First, we present a formulation of the Rodeo algorithm employing a general $d$-level ancilla qudit. Second, we introduce the concept of the \emph{Rodeo kernel}, defined as a two-frequency interferometer, which acts as a spectral filter in the energy domain. Finally, we propose a microcanonical protocol for the Rodeo algorithm. This protocol enables the estimation of entropic quantities through a single energy sweep and admits a natural interpretation as a Gaussian convolution of the density of states. To support the theoretical analysis, we perform numerical evaluations of the corresponding quantum circuit using ancilla qudits of dimensions three, four, and five. The simulations are performed for the one-dimensional Ising model, considering both spin-$\sfrac{1}{2}$ and spin-$1$ particles. The ancilla qutrit implementation exhibits an $18\%$ reduction in fluctuations compared to the qubit implementation. Our results show that qudits provide a framework for spectral analysis and thermodynamic characterization of multi-level quantum systems.
\end{abstract}


\maketitle
\section{Introduction}

In the early 1980s, Richard Feynman argued that quantum computers can achieve exponential speedups for certain problems by exploiting the exponentially large structure of Hilbert space~\cite{Feynman1982}. Shortly thereafter, David Deutsch formalized the notion of a universal quantum computer and demonstrated that classical devices cannot efficiently simulate general quantum systems, thereby establishing the distinctive computational power of quantum mechanics~\cite{Deutsch1985}.

Within this framework, the qudit—a multi-level quantum computational unit~\cite{Rains1999,Gottesman1999,Zhou2003}—can surpass the conventional two-level qubit for two main reasons: It provides a natural platform for simulating multi-level quantum systems and can enhance algorithmic efficiency. In particular, qudits enable an intrinsically parallel architecture by allowing the simultaneous implementation of multiple control operations, while offering an enlarged state space for information storage and processing. These features can reduce circuit complexity. Although maintaining high fidelity in multi-tone control is challenging, experimental realization on $5$- and $8$-level qudits in a trapped-ion system demonstrates competitive performance in implementing Grover's algorithm~\cite{Shi2026}. Coherent control of  $10$-level qudits has been demonstrated in optical platforms~\cite{Kues2017}. Furthermore, solid-state devices also constitute a hardware platform for the implementation of multi-level qudit systems~\cite{Nguyen2024}.

\begin{figure}[htbp]
    \centering
    \begin{minipage}{0.45\textwidth}
        \centering
        \begin{tabular}{c}
        \includegraphics[width=\textwidth]{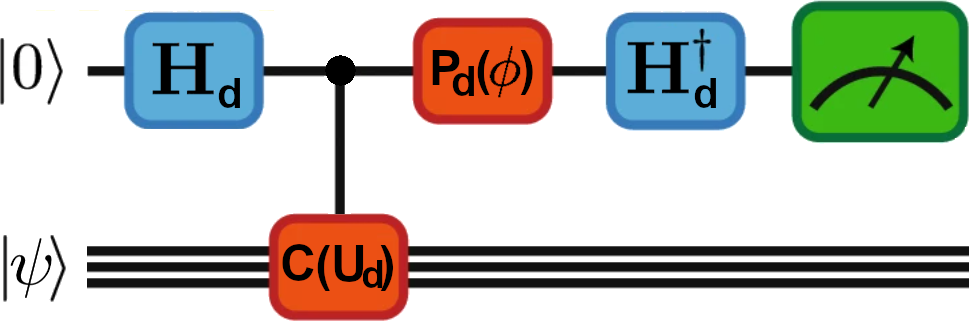} \\ (a)
        \end{tabular}
    \end{minipage}
    
    \vspace{1cm} 
    
    \begin{minipage}{0.475\textwidth}
        \centering
        \begin{tabular}{c}
        \includegraphics[width=\textwidth]{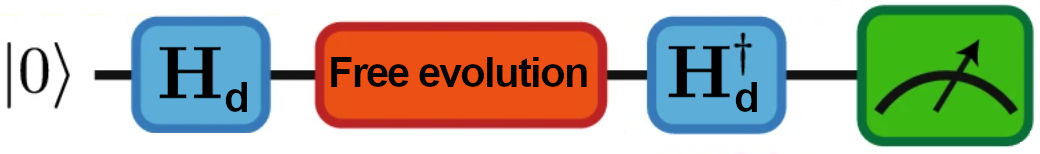} \\ (b)
        \end{tabular}
    \end{minipage}
    \hfill 
    \begin{minipage}{0.425\textwidth}
        \centering
        \begin{tabular}{c}
        \includegraphics[width=\textwidth]{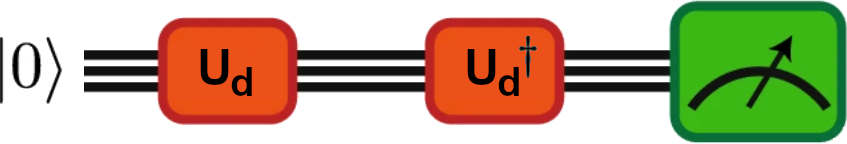} \\ (c)
        \end{tabular}
    \end{minipage}
\caption{Circuit diagrams illustrating (a) the Rodeo algorithm, (b) a Ramsey interferometer, and (c) a Loschmidt-echo protocol. In the Rodeo implementation, the free-evolution arm of the Ramsey interferometer is replaced by a controlled time-evolution operator followed by a phase shift. As the phase shift constitutes an attempt to reverse the effect of the controlled evolution on the ancilla qudit, the Rodeo circuit admits a natural interpretation in terms of a Loschmidt echo.}\label{fig:circuitRodeo}
\end{figure}

Inspired by recent implementations of Ramsey interferometry in multi-level quantum systems~\cite{Godfrin2018,Ilikj2025}, and motivated by its close conceptual relationship with the Rodeo algorithm~\cite{Choi2021,Qian2024,Gomes2025}, we formulate the Rodeo algorithm within a general $d$-dimension qudit framework. In Ramsey spectroscopy, a coherent superposition of internal states accumulates relative phases during a period of free evolution. In the Rodeo algorithm, this free-evolution arm is effectively realized by a controlled operator followed by a phase shift, see Fig.~\ref{fig:circuitRodeo}. In both settings, the $d$-dimensional quantum Fourier transform (QFT)~\cite{Stroud2002} naturally generalizes the role of the Hadamard gate used in qubit-based architecture. Beyond that, the Rodeo protocol requires access to a sufficiently general set of multi-level quantum gates~\cite{Brylinski2002,Brennen2005,Goss2022,Randall_2015,Gokhale_2019}. The proposed approach appears feasible to implement, given the availability of the required tools.

In the Rodeo protocol, the phase shift is introduced as a tentative attempt to invert the effect of the controlled operation on the ancilla qudit, thereby allowing a natural interpretation in terms of the Loschmidt echo. In this context, if quantum dynamics were perfectly reversible, a quantum state evolved forward in time would return exactly to its initial configuration upon a time-reversal operation. In simple words, the Loschmidt echo quantifies the survival ratio of a state under forward and backward evolution. Importantly, beyond serving as a measure of reversibility, the temporal dependence of the Loschmidt echo encodes the energy distribution associated with the initial state. In the Rodeo algorithm, this connection underlies the extraction of spectral information from time-domain interference measurements. 

We also introduce a single-state protocol that enables estimation of the density of states and microcanonical entropy using the Rodeo algorithm. This protocol is based on the convolution of a Fourier series with a Gaussian distribution. The resulting signal serves as a Gaussian smoothing filter applied to the quantum spectrum.

This paper is organized as follows. In Sec.~\ref{sec:theory}, we present the theoretical framework of the algorithm. In particular, Sec.~\ref{sec:revRodeo} revisits the Rodeo algorithm and describes its implementation in qudit systems through the step-by-step evolution of the joint ancilla–system state. In Sec.~\ref{sec:probResol}, we show how the width of the success-probability peak decreases as the ancilla dimension increases. In Sec.~\ref{sec:SA}, we derive the Rodeo kernel and show that it exhibits a small-amplitude, high-frequency harmonic perturbation that can improve spectral resolution, with this effect being more pronounced in the ancilla-qutrit implementation. In this section, we also discuss the connection with the Loschmidt amplitude. In Sec.~\ref{sec:entropic} we introduce the microcanonical ensemble protocol for entropy estimation, which interprets the Rodeo algorithm as a convolution-based spectral smoothing filter. Sec.~\ref{sec:numerical} presents the numerical evaluation of the proposed approach for the one-dimensional Ising model. Specifically, in Sec.~\ref{sec:spinHalf}, we consider spin-$\sfrac12$ particles and compare the qubit and qudit implementations. Then empirically demonstrate that the harmonic perturbation of the Rodeo kernel improves spectral resolution and reduces noise, with numerical results in good agreement with theoretical predictions. In Sec.~\ref{sec.spin1}, we study the spin-$1$ Ising model and show that the microcanonical protocol accurately reproduces the quantum degeneracy of the system. Finally, the conclusions and perspectives are presented in Sec.~\ref{sec:conclusion}.

\section{Theoretical Framework}
\label{sec:theory}
We initiate the formal treatment by revisiting the Rodeo algorithm and explicitly extending its framework to qudit architectures through a detailed derivation of the evolution of the joint ancilla-system quantum state across the full quantum circuit. 
\subsection{Revisiting the Rodeo Algorithm}
\label{sec:revRodeo}
\subsubsection{Initialization}

Let $\mathcal{H}$ be a finite-dimensional Hamiltonian with spectral decomposition
\begin{equation}
\mathcal{H}\ket{x} = E_x \ket{x},
\end{equation}
where $\{\ket{x}\}$ denotes the complete set of orthonormal energy eigenstates. 
Each eigenstate $\ket{x}$ is encoded in a tensor product of $k$ qudits, sufficient to represent any physically accessible state of the system under consideration. 
For a system composed of $d'$-level particles, $d'$-dimensional qudits constitute a natural — though not unique — choice of encoding. Under this correspondence, the number of qudits is equal to the number of particles, N. Hence, in the following, we set $k=N$. A general pure system state can therefore be expressed as
\begin{equation}
\ket{\psi} = \sum_{x=0}^{d'^N-1} c_x \ket{x}, 
\ \text{ where } \ 
c_x = \braket{x}{\psi}.
\end{equation}
For notational simplicity, the summation limits of the system basis states will be omitted in the following expressions.

We introduce an ancilla qudit of dimension $d$, with computational basis
$\{\ket{\ell}\}_{\ell=0}^{d-1}$. In general, for heterogeneous hardware architectures, the dimensions $d$ and $d'$ need not coincide. 
Assuming that the ancilla is initialized in the state $\ket{0}$, the joint initial state is given by
\begin{equation}
\ket{\Psi_0} = \ket{0} \otimes \ket{\psi} =  \sum_x c_x \ket{0}\otimes \ket{x}.
\end{equation}

The first step is to prepare the ancilla qudit in the uniform superposition state
\begin{equation}
\ket{+}_d =
\frac{1}{\sqrt{d}}
\sum_{\ell=0}^{d-1} \ket{\ell},
\end{equation}
which can be done by applying the $d$-dimensional quantum Fourier transform (QFT)~\cite{Cao2011} to the state $\ket{0}$, namely $\ket{+}_d = \mathbf{H}_d \ket{0}$, where
\begin{equation}
\mathbf{H}_d
=
\frac{1}{\sqrt{d}}
\sum_{\ell,n=0}^{d-1}
\exp(\frac{i 2\pi \ell n}{d})\ketbra{\ell}{n},
\label{eq:QFT}
\end{equation}
The resulting joint state is 
\begin{align}
\ket{\Psi_1}
& =\ket{+}_d \otimes \ket{\psi} = (\mathbf{H}_d\otimes \mathds{1}_{d'}^{\otimes N})  \ket{\Psi_0}, \\
&= \frac{1}{\sqrt{d}} \sum_{\ell=0}^{d-1} \sum_x c_x\, \ket{\ell} \otimes \ket{x},
\end{align}
where
\begin{equation}
    \mathds{1}_{d^{'}} = \sum_{q=0}^{d'-1} \ketbra{q}{q},
\end{equation}
denotes the identity operator acting on a single system qudit. Since $\ket{x}$ is a tensor product of $N$ qudits then $\mathds{1}_{d^{'}}^{\otimes N} \ket{x} = \ket{x} $.

\subsubsection{Loschmidt Echo-Like Circuit}
Following the Rodeo circuit, we apply a controlled unitary operation, defined as
\begin{equation}
C(U_{d'}) =
\sum_{n=0}^{d-1} \ketbra{n}{n} \otimes (U_{d'})^{n}.
\end{equation}
Although the Rodeo algorithm applies to arbitrary unitary observables, here we focus on solving the time-independent Schrödinger equation. Accordingly, we choose $U_{d'}$ to be the time-evolution operator
\begin{equation}
U_{d'} = e^{-i \mathcal{H} t}.
\label{eq:timeOperator}
\end{equation}
We adopt units such that $\hbar = 1$.  In the original formulation of the Rodeo algorithm, the evolution time $t$ is sampled from a Gaussian distribution. Nevertheless, alternative distributions are possible, including those defined by adaptive parametrization schemes. Recent results indicate that sampling $t$ according to a geometric series leads to improved sampling performance~\cite{Cohen2023,patkowski2026}. Applying the controlled evolution to the joint state $\ket{\Psi_1}$ yields
\begin{align}
\ket{\Psi_2}
&= C(U_{d'})\ket{\Psi_1} \nonumber \\
&=
\frac{1}{\sqrt{d}}
\sum_{\ell=0}^{d-1}
\sum_x
c_x\,
e^{-i E_x t \ell}
\ket{\ell} \otimes \ket{x},
\end{align}
where we have used the orthonormality of the computational basis. The next step is to apply a phase-shift operation to the ancilla qudit. The phase shift operator is
\begin{equation}
\mathcal{P}_d(\phi)
=
\sum_{n=0}^{d-1}
\exp(\frac{i 2\pi n \phi}{\pi d})
\ketbra{n}{n},
\end{equation}
as defined in Ref.~\cite{pudda2024}.
When the system state coincides with an eigenstate of the Hamiltonian, $\ket{\psi} = \ket{x}$, an appropriate choice of the phase parameter $\phi$ can reverse the effect of the controlled time-evolution operator. In practice, this phase is chosen as a trial value, which we parametrize as
\begin{equation}
\phi = \frac{Ed}{2}t,
\end{equation}
where $E$ is interpreted as a trial eigenvalue associated with the target eigenstate, and $t$ denotes the evolution time defined in eqn.~(\ref{eq:timeOperator}).

Since the phase shift acts exclusively on the ancilla qudit, the resulting joint state is
\begin{align}
\ket{\Psi_3}
&=
\left[
\mathcal{P}_d(\phi) \otimes \mathds{1}_{d^{'}}^{\otimes N}
\right]
\ket{\Psi_2}
\nonumber \\
&=
\frac{1}{\sqrt{d}}
\sum_{\ell=0}^{d-1}
\sum_x
c_x\,
e^{-i \omega_x t \ell}
\ket{\ell} \otimes \ket{x},
\label{psi3}
\end{align}
where $\omega_x = E_x - E $.

\subsubsection{Final State}

The next step of the circuit consists of a basis transformation on the ancilla qudit, returning it from the Fourier basis to the computational basis. This is achieved by applying the inverse quantum Fourier transform,
\begin{equation}
\mathbf{H}_d^{\dagger}
=
\frac{1}{\sqrt{d}}
\sum_{n,\ell'=0}^{d-1}
\exp(- \frac{i 2\pi n \ell'}{d} )
\ketbra{n}{\ell'}.
\label{eq:QFT_dagger}
\end{equation}
The resulting joint state is therefore
\begin{align}
\ket{\Psi_4}
&=
\left(
\mathbf{H}_d^{\dagger} \otimes \mathds{1}_{d^{'}}^{\otimes N}
\right)
\ket{\Psi_3}
\nonumber \\
&=
\frac{1}{d}
\sum_{n,\ell=0}^{d-1}
\sum_x
c_x\,
e^{-i\left( \omega_xt + \frac{2\pi}{d} n\right)\ell}
\ket{n} \otimes \ket{x}.
\label{psi4}
\end{align}

Finally, the algorithm terminates by measuring the ancilla qudit. Before discussing this measurement step, we examine the projections of the final joint quantum state.

\subsubsection{Final State Projections}

Projecting the final joint state $\ket{\Psi_4}$ onto 
$\ket{n} \otimes \ket{\psi}$ yields the amplitude
\begin{align}
A(n) &=(\bra{n}\otimes\bra{\psi})\ket{\Psi_4} \nonumber\\
&=
\frac{1}{d}
\sum_{\ell=0}^{d-1}
\sum_x
|c_x|^2
e^{-i\left(\omega_x t + \frac{2\pi n}{d}\right)\ell}.
\label{eq:projAmpl}
\end{align}
The sum in $\ell$ can be recognized as a discrete Fourier transform, which leads to the Dirichlet kernel~\cite{He2019}, as demonstrated in Appendix~{\ref{amplitude}}, then
\begin{equation}
A(n)
=
\frac{e^{i \frac{\pi n}{d}}}{d}
\sum_x
|c_x|^2
e^{-i \omega_x' \frac{t}{2}}
\frac{\sin \left(\omega_x \frac{t d}{2} \right)}
{\sin \left( \omega_x \frac{t}{2} + \frac{\pi n}{ d} \right)},
\label{eq:dirchlet_kernel}
\end{equation}
where $\omega_x' = (d-1)\omega_x$.
The corresponding probability distribution for measuring the ancilla qudit in the $n$-th level and the system in its initial state $|\psi\rangle$ is therefore
\begin{align}
P_d(n,\psi)
&= |A(n)|^2, \nonumber \\
&=
\frac{1}{d^2}
\left|
\sum_x
|c_x|^2
e^{-i \omega_x' \frac{t}{2}}
\frac{\sin \left(\omega_x \frac{t d}{2} \right)}
{\sin \left(\omega_x \frac{t}{2}+ \frac{\pi n}{ d} \right)}
\right|^2.
\end{align}

\subsubsection{Success Probability}
\label{sec:probResol}
In its original formulation, the Rodeo algorithm requires a hypothesized solution to the eigenvalue problem as input. A key requirement of this approach, as emphasized by the original authors, is that the input state exhibits substantial overlap with the target eigenstate of the Hamiltonian. Several strategies are proposed for constructing such an initial state, including preparation guided by physical intuition regarding the structure of the system’s wave functions~\cite{Choi2021}.

In the idealized case where the input state closely approximates an eigenstate, $\ket{\psi} \simeq \ket{x}$, the overlap coefficients satisfy
\(
|c_{x'}|^2 \approx \delta(x,x').
\)
Under this assumption, the probability distribution for the ancilla measurement outcome simplifies to
\begin{equation}
P_d(n) \approx \frac{1}{d^2}
\left|
\frac{
\sin \left(\omega_x \frac{td}{2}\right)
}{
\sin \left(\omega_x \frac{t}{2} + \frac{\pi n}{d}\right)
}
\right|^2.
\label{eq:p_dn_x}
\end{equation}

Successful solution of the eigenproblem corresponds to measuring the ancilla in the state $\ket{n} = \ket{0}$. The associated success probability is therefore
\begin{equation}
P_d(0) \approx \frac{1}{d^2}
\left|
\frac{
\sin \left(\omega_x \frac{t d}{2}\right)
}{
\sin \left(\omega_x \frac{t}{2} \right)
}
\right|^2.
\end{equation}

In the qubit case, corresponding to $d = 2$, this expression reduces to
\begin{equation}
P_2(0) \approx   \cos^2 \left[(E_x - E)\frac{t}{2}\right],
\end{equation}
where we have used the trigonometric identity $\sin(2\theta) = 2 \cos(\theta)\sin(\theta)$. This result is in direct agreement with the success probability obtained in the original formulation of the Rodeo algorithm. 

\begin{figure}[!ht]
\begin{tabular}{c c}
\includegraphics[scale=0.22] {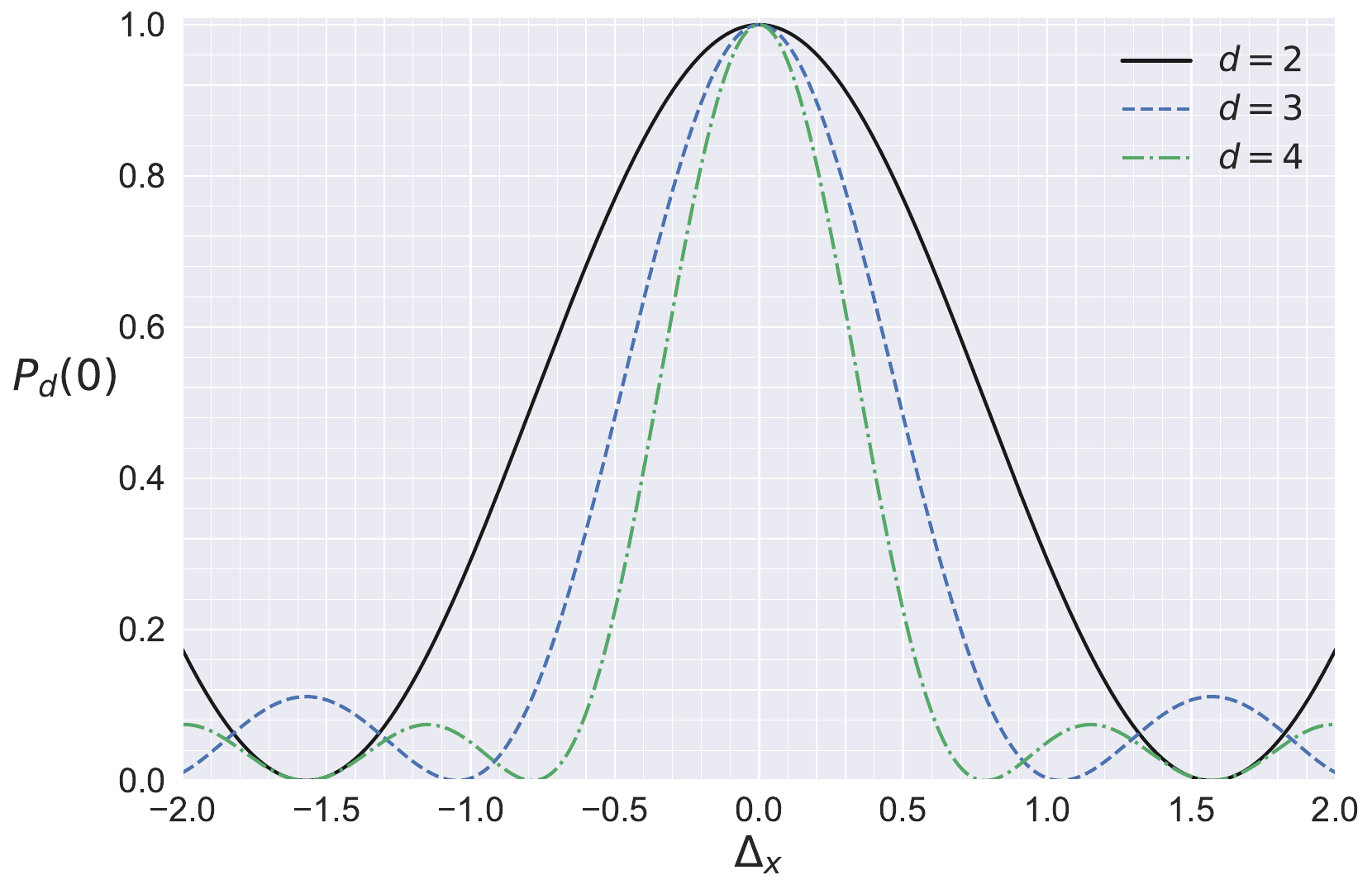} & 
\includegraphics[scale=0.22]{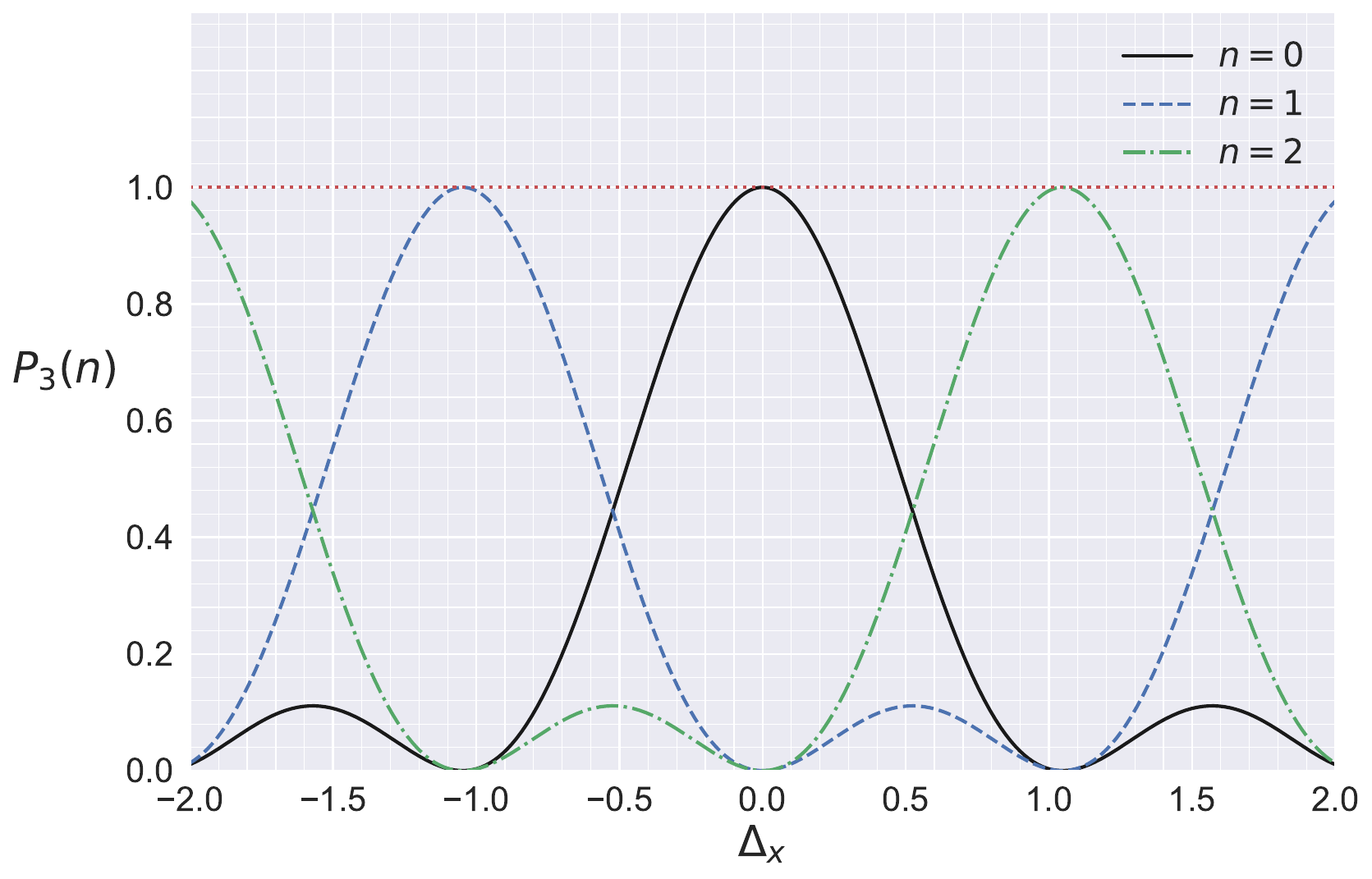} \\ (a) & (b)
\end{tabular}
\caption{Probability of measuring a $d$-level ancilla qudit in the $n$th state when the system is prepared in a Hamiltonian eigenstate [see eqn.~(\ref{eq:p_dn_x})], where here $\Delta_x  = \omega_xt/2$. (a) Probability distribution for $n = 0$ and $d = 2, 3,$ and $4$. (b) Probability distribution for $d = 3$ and $n = 0, 1,$ and $2$; the dotted red curve corresponds to the sum over all $n$. }
\label{fig:probability}
\end{figure}

The behavior of eqn.~(\ref{eq:p_dn_x}) is illustrated in Fig.~\ref{fig:probability}. Panel (a) displays the success probability for different ancilla dimensions:  the qubit case ($d=2$) is represented by the solid black curve, the qutrit  ($d=3$) by the dashed blue curve, and the ququart ($d=4$) by the dash-dotted green curve. Panel (b) shows the probability distribution for measuring the ancilla qutrit in its different levels. The probability of observing the state $n=0$ is shown by the solid black curve, $n=1$ by the dashed blue curve, and $n=2$ by the dash-dotted green curve. The dotted red curve corresponds to the sum over all values of $n$, demonstrating the normalization of the probability distribution.

As the qudit dimension increases, the greater diversity of possible outcomes leads to a progressively narrower distribution of success probabilities. In state preparation, lower success probabilities increase the number of circuit executions required to prepare the target state, thereby increasing the post-selection overhead.  The width of the success-probability distribution can be controlled by adjusting the parameters of the probability distribution from which the evolution times are sampled. In the following section, we investigate the influence of the temporal distribution and demonstrate that the standard deviation of the Gaussian distribution governs the width of the Rodeo response. In the small-detuning regime, the sensitivity of the qudit implementation to deviations from resonance can be matched to that of the qubit implementation by choosing
\begin{equation}
\sigma_d = \sigma_2 \sqrt{\frac{3}{d^2-1}},
\label{eq:ratio_sigma}
\end{equation}
where $\sigma_d$ and $\sigma_2$ denote the standard deviations of the Gaussian distributions for ancilla dimensions $d$ and $2$, respectively.

Furthermore, outcomes with \(n\neq 0\) need not be discarded as uninformative shots. According to eqn.~(\ref{eq:p_dn_x}), \(P_d(n)\) exhibits maxima at
\begin{equation}
\omega_x t = -\frac{2\pi n}{d}
\pmod{2\pi}.
\label{eq:max_pdn}
\end{equation}
Thus, nonzero outcomes provide information about the detuning, including its sign, see Fig.~\ref{fig:probability}~(b). This information could be used to adaptively update the trial energy \(E\), potentially improving the efficiency of energy estimation compared with strategies that rely exclusively on successful outcomes.  We derive eqns.~(\ref{eq:ratio_sigma}) and (\ref{eq:max_pdn}) in Appendix~\ref{sec:success}.

By contrast, expectation values may be less sensitive to the redistribution of probabilities among individual outcomes because they involve contributions from multiple outcomes. This possibility is discussed further below.

\subsection{Statistical Procedures}
\label{sec:SA}
Within the Copenhagen interpretation of quantum mechanics, physical observables are not assigned definite values prior to measurement; rather, only probabilistic measurement outcomes are meaningful. Accordingly, the relevant quantities to analyze in the Rodeo algorithm are its measurement statistics.

\subsubsection{$Z_d$ Expectation Value}

The measurement is implemented by projecting the final quantum state onto the observable \( Z_d \), commonly referred to as the \emph{clock operator}, whose eigenbasis defines the computational measurement basis. By definition~\cite{pudda2024},
\begin{equation}
Z_d
=
\sum_{n=0}^{d-1}
\omega^n \ketbra{n}{n}
=
\sum_{n=0}^{d-1}
e^{i \frac{2\pi n}{d}}
\ketbra{n}{n}.
\end{equation}
The clock operator can equivalently be expressed as
\begin{equation}
Z_d = \mathbf{H}_d X_d \mathbf{H}_d^{\dagger},
\end{equation}
where
\begin{equation}
X_d = \sum_{\ell=0}^{d-1} \ketbra{(\ell + 1)\,\mathrm{mod}\, d}{\ell},
\end{equation}
denotes the shift operator, satisfying \( X_d \ket{n} = \ket{n+1} \). 
The modular operation ensures that \( X_d \ket{d-1} = \ket{0} \), thereby imposing periodic boundary conditions on the finite-dimensional computational basis. This condition is important for further discussion.

We interpret the measurement outcome as the ancilla qudit level, $n=0,1,\cdots,d-1$, thereby enabling direct determination of the coefficients of the $Z_d$ operator. Thus, we are effectively measuring the clock operator eigenvalues. The corresponding measurement signal is defined as the expectation value
\begin{equation}
  h_d(E,\psi \,|\, t) =  \langle Z_d\otimes\mathds{1}_{d^{'}}^{\otimes N}\rangle = \bra{\Psi_4}(Z_d\otimes\mathds{1}_{d^{'}}^{\otimes N}) \ket{\Psi_4},
\end{equation}
which explicitly evaluates to
\begin{equation}
h_d(E,\psi \,|\, t)
=
\frac{1}{d^2}
\sum_{\ell,p,n=0}^{d-1}
\sum_x |c_x|^2
e^{-i\left[\omega_x t(\ell - p) + \frac{2\pi n}{d}(\ell - p - 1)\right]}.
\label{eq:he_total}
\end{equation}
The sum over $n$ is identified as the discrete Fourier transform over the finite cyclic group of order $d$, yielding
\begin{equation}
    \sum_{n=0}^{d-1} e^{-i\frac{2\pi n}{d}(\ell - p - 1)}
    =
    d\,\delta_{\ell,\,s},
\end{equation}
where $s = (p+1)\bmod d$, then eqn.~(\ref{eq:he_total}) can be written as
\begin{equation}
h_d(E,\psi \,|\, t)
=\frac{1}{d}
\sum_{\ell,p=0}^{d-1}
\sum_x
|c_x|^2
e^{-i\left[\omega_x t(\ell - p)\right]}\delta_{\ell,\,s}.
\label{eq:he_total_eval}
\end{equation}
The Kronecker delta, $\delta_{\ell, \,s}$, constrains the index pairs $(\ell,p)$ to satisfy $\ell = (p+1)\ \mathrm{mod}\ d$. Thus, for $p = 0, 1, \cdots, d-2$, the only surviving contributions to eqn.~(\ref{eq:he_total_eval}) arise from index pairs satisfying $\ell - p = 1$. 
There are $(d-1)$ such pairs, each contributing a term proportional to 
$e^{-i\omega_x t}$. The remaining contribution originates from the periodic boundary condition 
of the finite-dimensional basis, corresponding to the pair 
$p = d-1$ and $\ell = 0$. In this case, $\ell - p = -(d-1)$, 
which yields a term proportional to 
$e^{+i(d-1)\omega_x t}$. Collecting all $d$ contributions, we can define the Rodeo kernel as
\begin{equation}
    K_d(\omega_x , t)  = \frac{d-1}{d} e^{-i\omega_x t} + \frac{1}{d} e^{+i\omega_x' t},
\end{equation}
recalling that $\omega_x' = (d-1)\omega_x$.
Therefore, eqn.~(\ref{eq:he_total_eval}) takes the exact form
\begin{equation}
    h_d(E,\psi \,|\, t) =
    \sum_x |c_x|^2 K_d(\omega_x, t).
\label{H_echo}
\end{equation}

It is worth noting that, if the input system state is an eigenstate of the Hamiltonian, $\ket{\psi} =\ket{x}$, just one coefficient survives in eqn.~(\ref{H_echo}), then  $h_d(E,x \,|\, t) = K_d(\omega_x , t)$. Moreover, the Rodeo kernel can be seen as a two-frequency interferometer: a high-amplitude, low-frequency signal and a low-amplitude, high-frequency signal. The dependence of the expectation value on the ancilla dimension arises from interference between the two signals. In the ancilla qubit implementation, $d = 2$, these two terms combine to give
\begin{equation}
    h_2(E,\psi \,|\, t) = \sum_x |c_x|^2 \cos\!\left[(E_x - E)t\right],
    \label{H_echo_qubit}
\end{equation}
in agreement with the existing literature~\cite{Gomes2025, Rocha024}. In this situation, the signal contains a single-frequency component; therefore, no interference occurs. Consequently, interference effects first appear in the qutrit implementation ($d=3$). Moreover, for $d=3$, the higher-frequency term lies closer to resonance and has a larger amplitude than for $d>3$. So the interplay between the two frequency components is expected to be most pronounced in the qutrit implementation.

Alternatively, in terms of the Loschmidt amplitude
\begin{equation}
L(t) = \sum_x |c_x|^2 e^{-i (E_x - E) t},
\end{equation}
eqn.~(\ref{H_echo}) can be expressed as
\begin{equation}
    h_d(E,\psi \,|\, t)
    =
    \frac{d-1}{d}\,L(t)
    +
    \frac{1}{d}\,L^*((d-1)t),
    \label{eq:h_L_exact}
\end{equation}
where $L^*$ denotes the complex conjugate of
the Loschmidt amplitude. The second term constitutes a harmonic contamination of
weight $1/d$ that vanishes in the limit $d\to\infty$, recovering $h \to L(t)$
exactly, thereby corroborating the interpretation of the Rodeo algorithm as a realization of a Loschmidt echo\footnote{Zeros of the Loschmidt amplitude correspond to nonanalyticities in the dynamical free energy, signaling dynamical quantum phase transitions (DQPTs) in the thermodynamic limit.}. 


\subsubsection{Ensemble Average}
\label{sec:Ensemble}

We define the \emph{spectral amplitude} (SA) of the Rodeo algorithm as the ensemble average of the expectation value of the operator $Z_d \otimes \mathds{1}_{d'}^{\otimes N}$.
It is the mean value of $h_d(E,\psi \,|\, t)$ over repeated realizations of the algorithm, where the
evolution time $t$ is randomly sampled from a probability density function
(PDF) denoted by $P(t \,|\, \{X\})$. Here, $\{X\}$ represents the set of parameters characterizing the PDF. The SA is therefore quantified by
\begin{equation}
G_d(E,\psi)=
\int_{-\infty}^{\infty}
h_d(E,\psi \,|\, t)\,
P(t \,|\, \{X\})\,
\mathrm{d}t.
\label{eq:SA_definition}
\end{equation}

Following the original formulation of the Rodeo algorithm, we choose the PDF
to be a normal (Gaussian) distribution,
\begin{equation}
P(t \,|\, \mu,\sigma)
=
\frac{1}{\sigma\sqrt{2\pi}}
e^{-\frac{(t-\mu)^2}{2\sigma^2}},
\label{eq:Gaussian_PDF}
\end{equation}
where the set of parameters {$\{X\}=\{\mu,\sigma\}$}, denotes the mean and standard deviation, respectively. The Gaussian distribution is relevant for specific interpretations of the algorithm considered in this work.
Substituting eqn.~(\ref{eq:Gaussian_PDF}) and (\ref{H_echo}) into
eqn.~(\ref{eq:SA_definition}) yields
\begin{equation}
G_d(E,\psi)= \sum_{x} |c_x|^2 \mathcal{T}_P(\omega_x)
\label{MedGauss}
\end{equation}
where
\begin{align} \nonumber
 \mathcal{T}_P(\omega_x) = &  \frac{d-1}{d} \left( \frac{1}{\sigma \sqrt{2\pi}}\int^{\infty}_{-\infty}  e^{-\frac{(t-\mu)^2}{2\sigma^2}} e^{-i\omega_x t} \mathrm{d}t\right) \\
  & +   \frac{1}{d}  \left(\frac{1}{\sigma \sqrt{2\pi}}\int^{\infty}_{-\infty}  e^{-\frac{(t-\mu)^2}{2\sigma^2}} e^{i\omega_x' t} \mathrm{d}t\right).
\end{align}
 These integrals correspond to the Fourier transform of a Gaussian and can be computed analytically:
\begin{equation}
    \frac{1}{\sigma \sqrt{2\pi}}\int^{\infty}_{-\infty} e^{\pm i\omega t} e^{-\frac{(t-\mu)^2}{2\sigma^2}} \mathrm{d}t  = e^{-\frac{\sigma^2\omega^2}{2}}e^{\pm i\omega \mu}.
\end{equation}
This leads to
\begin{equation}
    \mathcal{T}_P(\omega_x) =
    \frac{d-1}{d} e^{-\frac{\sigma^2 \omega_x^2}{2}}e^{-i \omega_x \mu} +
    \frac{1}{d} e^{-\frac{\sigma^2 \omega_x'^2}{2}}e^{+i \omega_x'  \mu } .
    \label{eq:SA_general}
\end{equation}

In our numerical evaluation, we set $\mu = 0$ to eliminate the
oscillatory phase factor and focus exclusively on the spectral filtering
properties of the SA. In this case, the expression simplifies to
\begin{align}  \nonumber
   \mathcal{T}_P(\omega_x)
     &=
    \frac{d-1}{d} e^{-\frac{\sigma^2 \omega_x^2}{2}}
    +
    \frac{1}{d} e^{-\frac{\sigma^2 {\omega'}_x^2}{2}}, \\
    & = e^{-\frac{\sigma^2 \omega_x^2}{2}} 
  \left[ 1 - \left( \frac{1 - e^{-\frac{\sigma^2 ({\omega_x'}^2 - \omega_x^2)}{2}} }{d} \right) \right].
    \label{eq:SA_gaussian}
\end{align}
Therefore, the resulting Rodeo response is governed by Gaussian profiles whose width can be tuned by appropriately choosing $\sigma$. Moreover, for finite $\omega_x$, the Gaussian damping in the second term is much stronger than that of the main term. Thus, the higher-frequency harmonic contamination decays rapidly, underscoring its perturbative nature.
For $d=2$, $\omega_x' = \omega_x$, so the perturbative term is canceled out, leading to 
\begin{equation}
    G_2(E,\psi) =
    \sum_x |c_x|^2\, e^{-\frac{\sigma^2 (E_x - E)^2}{2}},
    \label{eq:SA_qubit}
\end{equation}
corroborating our previous result~\cite{Gomes2025, Rocha024}, where we interpreted the SA as the probability of measuring the system state $\ket{\psi}$ with a given energy $E$. Besides, the limit $d\to \infty$ of eqn.~(\ref{eq:SA_gaussian}) recovers the result for the ancilla qubit implementation. These results therefore support the conjecture that the expectation values are only weakly affected by the ancilla qudit's dimension.

To create a metric for the perturbative effects, let us define the relative difference between the SA obtained using an ancilla qudit and that obtained with an ancilla qubit as
\begin{equation}
   \Delta_2 G_d(E,\psi) =  1 - \frac{G_d(E,\psi)}{G_2(E,\psi)}.
    \label{eq:relDiff}
\end{equation}
Since the Gaussian damping in $G_d(E,\psi)$ is stronger than in $G_2(E,\psi)$—that is, the former approaches zero faster than the latter—this relation can be considered well defined over the entire energy range.  

We aim to show that the Rodeo algorithm can be adapted into a quantum protocol for estimating the density of states (DoS), from which thermodynamic quantities such as the microcanonical entropy can be obtained directly. 

\subsection{Single-State Microcanonical Protocol}
\label{sec:entropic}

Recently, we proposed a protocol based on the Rodeo algorithm for estimating the entropy of a quantum system \cite{Rocha024}. The protocol consists of summing the SA over all computational basis states used to represent the system. A central limitation of this protocol is the exponential growth of the Hilbert-space dimension: for an $N$-particle system composed of $d'$-level subsystems, the computational basis contains $d'^N$ states. To mitigate this issue, we consider preparing the input state as a homogeneous superposition of the Hamiltonian eigenstates,
\begin{equation}
\ket{\psi_{\{x\}}}
= \frac{1}{\sqrt{d'^N}} \sum_x \ket{x},
\label{eq:homogenousInput}
\end{equation}
for which the overlap coefficients satisfy $|c_x|^2 = 1/d'^N$. For a homogeneous input state, which exhibits equal overlap with all energy eigenstates, the SA effectively reconstructs the DoS by naturally accounting for the contributions of degenerate eigenstates. This result can be established by considering the limit of eqn.~(\ref{eq:SA_gaussian}) as $\sigma \to \infty$:
\begin{equation}
\lim_{\sigma \to \infty} \mathcal{T}_P(\omega_x) = \delta(E_x, E).
\label{eq:limit_SA}
\end{equation}
Substituting eqn.~(\ref{eq:homogenousInput}) into eqn.~(\ref{MedGauss}) and taking the above limit, we obtain
\begin{align}
G_d(E,\psi_{\{x\}}) &:= g_d(E) \\
&\approx \frac{1}{d'^N} \sum_x \delta(E_x,E)
= \frac{\Omega(E)}{d'^N},
\label{eq:DoS_approach}
\end{align}
where $\Omega(E)$ denotes the number of states (NoS) at energy $E$.
This approach directly connects the SA with the DoS.
The microcanonical entropy,
\begin{equation}
S(E) = k_B \ln g(E) + C,
\end{equation}
where $k_B$ is the Boltzmann constant and $C = N k_B \ln d'$, then follows naturally\footnote{Since only entropy differences are thermodynamically relevant, the constant $C$ may be safely neglected.}. This formulation therefore provides a framework for studying quantum effects in thermodynamic systems~\cite{Goold2016}.

It is instructive to build an analogy with the canonical partition function. Eqn.~(\ref{eq:SA_qubit}) for the homogeneous input can be written as
\begin{align}
G_2(E,\psi_{\{x\}})
&= \frac{1}{d'^N}  \sum_x e^{-\frac{\sigma^2 (E_x - E)^2}{2}}, \nonumber \\
&= \frac{1}{d'^N}
\Tr \left[
e^{-\frac{\sigma^2}{2}(\mathcal{H}-E\mathds{1})^2}
\right].
\end{align}
On the other hand, the canonical partition function is given by
\begin{align}
Z(\beta)
&= \Tr \left(e^{-\beta \mathcal{H}}\right), \nonumber \\
&= \int \Omega(E)\, e^{-\beta E}\, \mathrm{d}E .
\end{align}
Then, by analogy, we may write
\begin{align}
\Tr \left[
e^{-\frac{\sigma^2}{2}(\mathcal{H}-E\mathds{1})^2}
\right]
&= \int \Omega(E')
e^{-\frac{\sigma^2}{2}(E'-E)^2}
\mathrm{d}E', \nonumber \\
&= \Big(\Omega \ast \mathcal{G}\Big)(E),
\label{eq:covOmega}
\end{align}
where $\mathcal{G}(E'-E)=\exp[-\sigma^2(E'-E)^2/2]$ is the Fourier transform of the Gaussian distribution. Since the Fourier transform of a Fourier series is a set of impulses, as $\Omega(E)=\sum_x \delta(E_x-E)$, then eqn.~(\ref{eq:SA_qubit}) follows directly from the convolution theorem,
\begin{equation}
    \big(u \ast v\big) = \mathcal{F}\{\mathcal{F}^{\dagger}\{u\}\mathcal{F}^{\dagger}\{v\}\},
\end{equation}
corroborating the analogy. Moreover, the convolution recovers $\Omega(E)$ in a smoothed form, $\Omega_{\sigma}(E)$, whose width is determined by $\sigma$. Consequently, the approach in eqn.~(\ref{eq:DoS_approach}) becomes the identity
\begin{equation}
g_d(E)
= \frac{\Omega_\sigma(E)}{d'^N},
\label{eq:SA_DOS}
\end{equation}
when $g_d(E)$ is interpreted as the smoothed density of states. This convolution structure reveals that the Rodeo algorithm acts as a spectral filter that applies Gaussian smoothing to the Hamiltonian spectrum.

This formulation, therefore, realizes a microcanonical ensemble protocol~\cite{gross2001microcanonical,Rocha2025}, in which the DoS is estimated from a single-state input given by eqn.~(\ref{eq:homogenousInput}). 
Although the protocol avoids explicit enumeration of the exponentially large computational basis, it remains statistically limited: small $g(E)$ values become difficult to distinguish from statistical fluctuations, as further discussed.

\section{Numerical Evaluation}
\label{sec:numerical}
In this section, we present a numerical analysis of the Rodeo algorithm based on a numerical linear-algebra evaluation of the joint ancilla-system state as it evolves through the Rodeo circuit. We apply the method to the one-dimensional Lenz-Ising model, described by the Hamiltonian
\begin{equation}
    \mathcal{H}_{d'} = -J \sum_{\langle i,j\rangle} S^z_i S^z_j ,
\end{equation}
where the symbol $\langle i,j\rangle$ represents the nearest-neighbor pairs along the chain, with $i = 0, \cdots, N-1$ and $j = i+1$, where $N$ is the number of spins in the chain. Periodic boundary conditions are assumed, so that site $N$ is identified with site $0$.  The parameter $J$ is the exchange coupling constant, and $S^z_{d', i}$ is the spin operator along the $z$ direction acting on the $i$-th particle in a $d'$-dimensional Hilbert space.  In our choice of units, energy is measured in units of $JS^2$, where $S$ denotes the spin magnitude.

In the graphs presented here, black dots denote the numerical evaluations, and error bars represent the associated standard deviations. Solid red curves represent theoretical predictions, and vertical blue lines, when present, indicate exact values.

\subsection{1D Spin $\frac{1}{2}$ Ising Model}
\label{sec:spinHalf}

\begin{figure}[!ht]
\begin{tabular}{c c}
\includegraphics[scale=0.22]{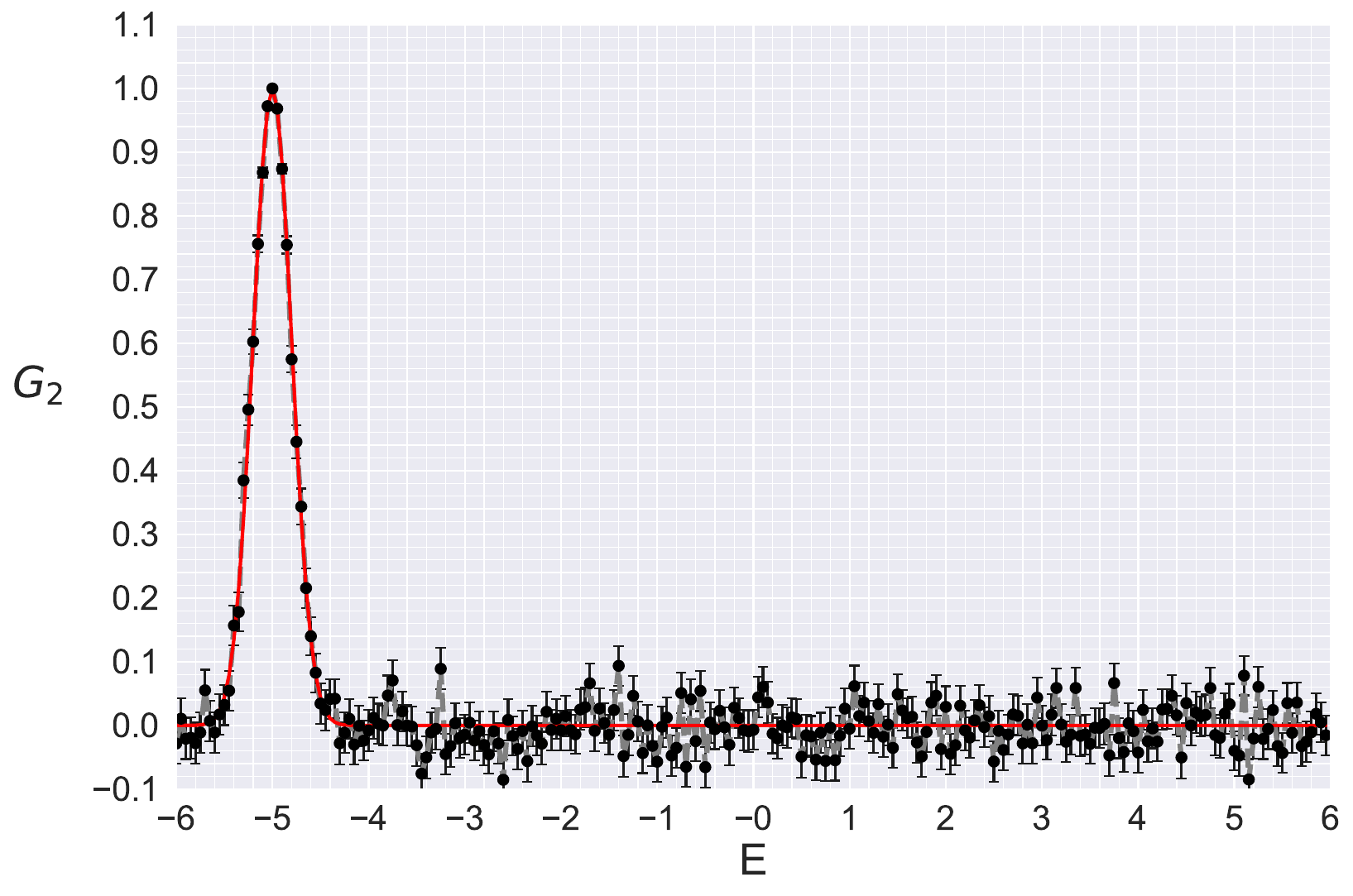} &
\includegraphics[scale=0.22]{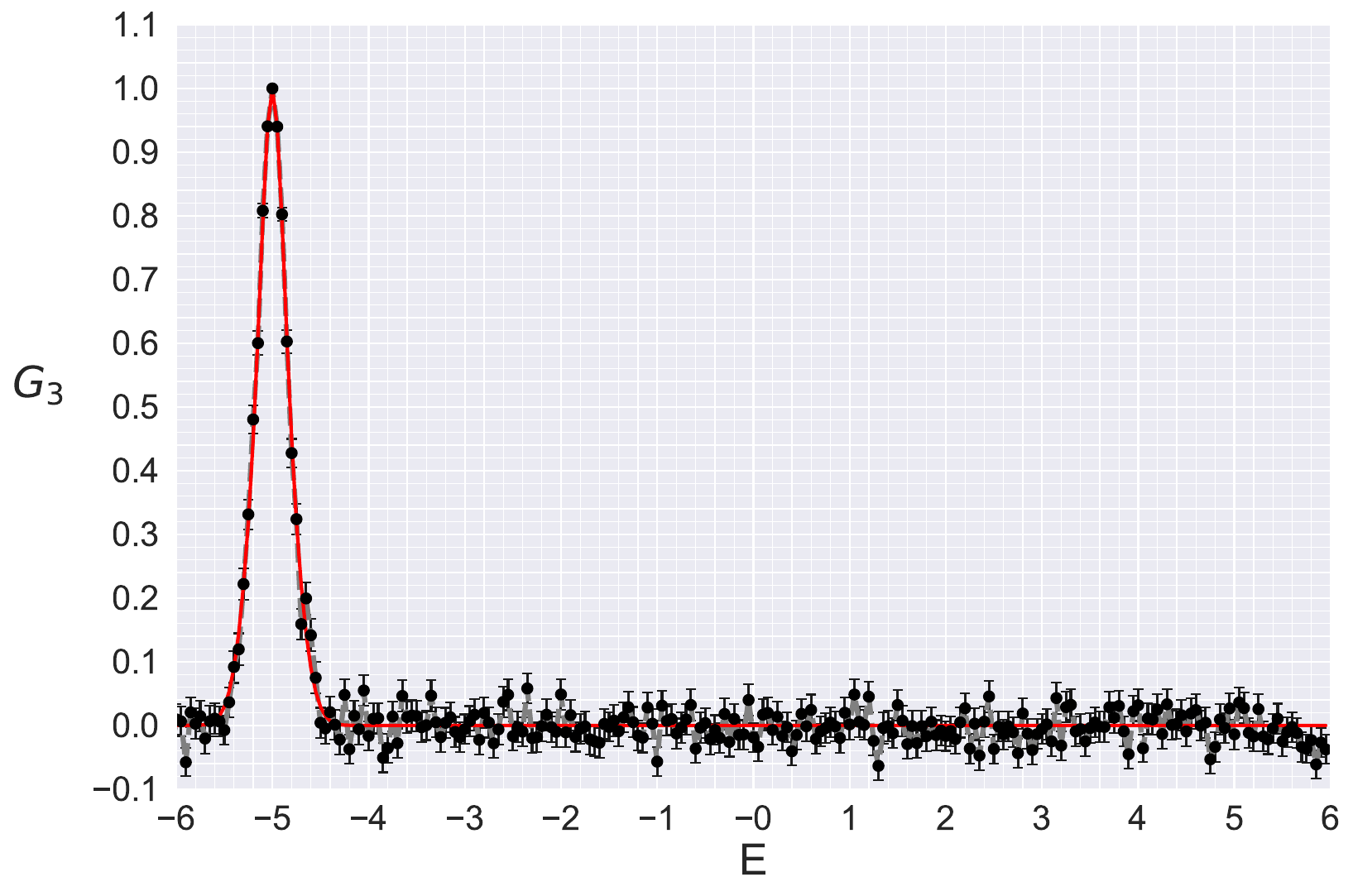} \\ 
(a) & (b) \\
\includegraphics[scale=0.22]{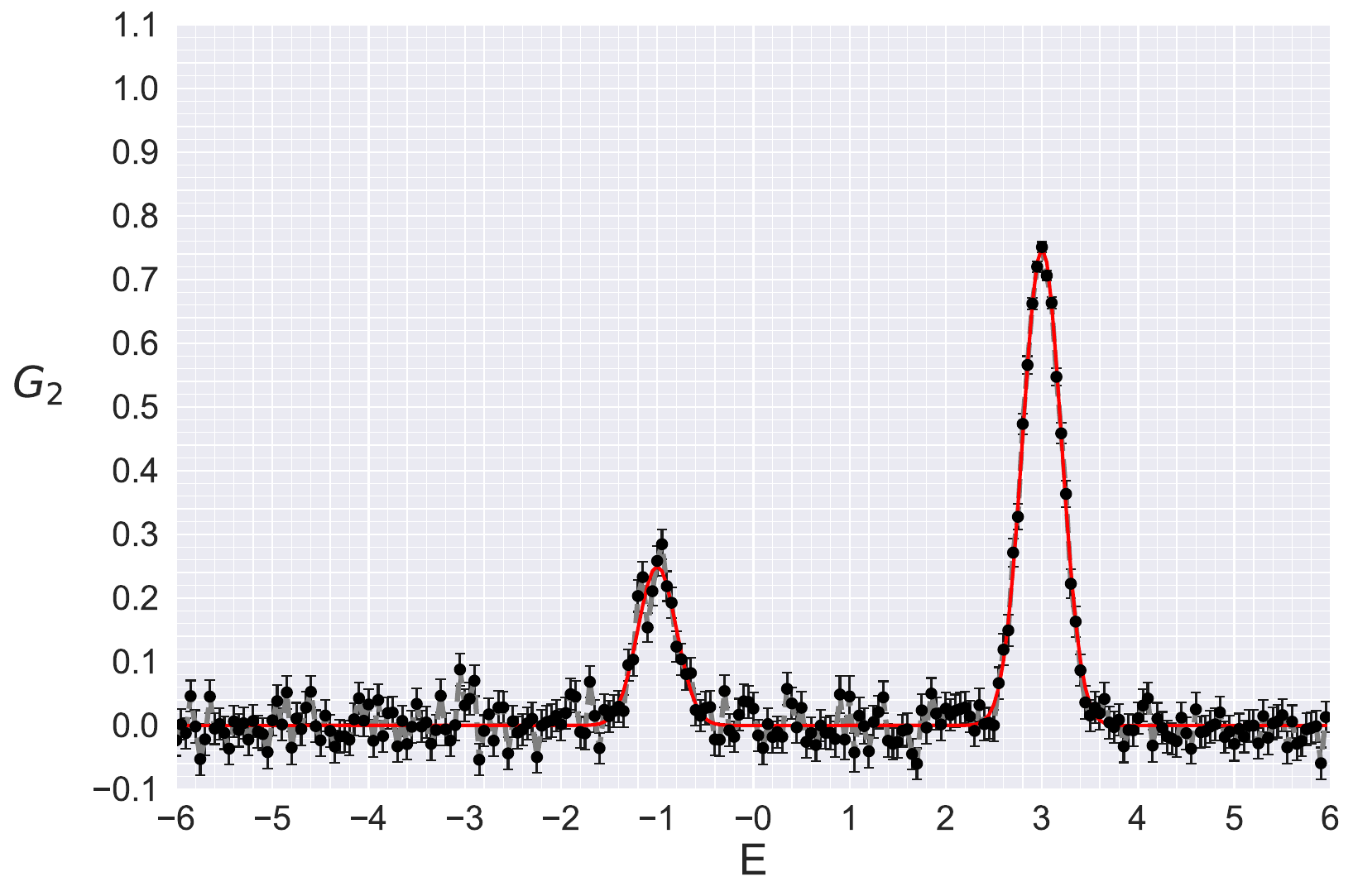} &
\includegraphics[scale=0.22]{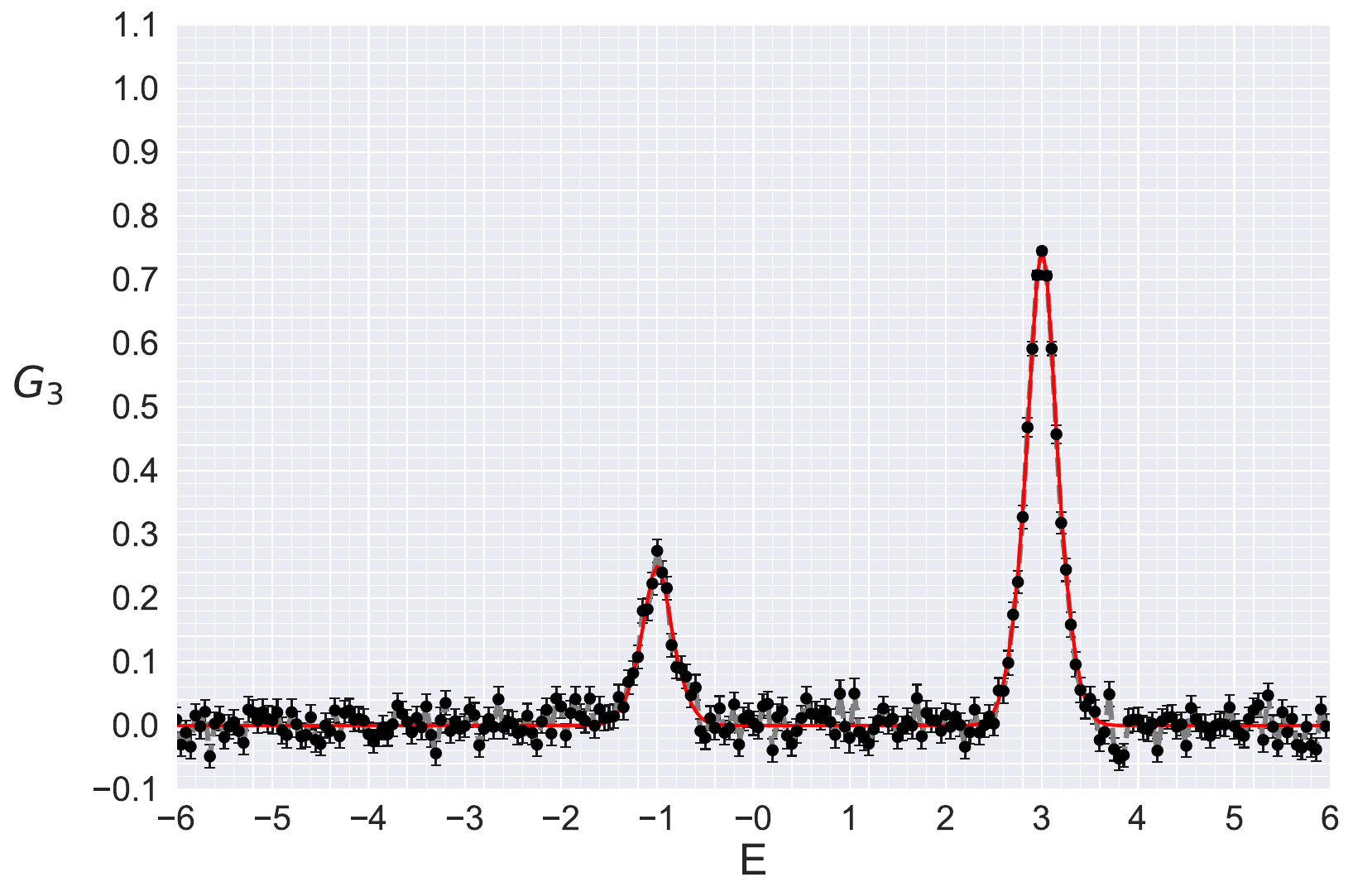} \\  (c) & (d)
\end{tabular}
\caption{Numerical evaluation of the Rodeo algorithm applied to the one-dimensional spin-$\sfrac{1}{2}$ Ising model with $N=5$ spins. The averages are over $500$ distinct evolution times per energy level, drawn from a Gaussian distribution with $\sigma = 5$. Panels (a) and (b) for system state $\ket{\psi} = \ket{0}$, while panels (c) and (d) $\ket{\psi} = \sfrac{1}{2}\ket{1} + \sfrac{\sqrt{3}}{2}\ket{5}$. In panels (a) and (c), the ancilla qubit ($d=2$) implementation, and (b) and (d) the ancilla qutrit ($d=3$) is employed. The local maxima of $G_d$ are interpreted as the probabilities of measuring the state $\ket{\psi}$ at energy $E$.}
\label{fig:IsingHalf}
\end{figure}

In Fig.~\ref{fig:IsingHalf} we present the results for the regular one-dimensional spin-$\sfrac{1}{2}$ Ising model. For this case ($d'=2$), $S_2^z$ corresponds to the Pauli-$Z$ matrix,
\begin{equation}
    S_2^z =
    \begin{pmatrix}
        1 & 0 \\
        0 & -1
    \end{pmatrix}
    = \ketbra{0}{0} - \ketbra{1}{1}.
\end{equation}
We consider a chain of \(N=5\) particles and sample \(N_t=500\) evolution times from a Gaussian distribution with standard deviation \(\sigma=5\). The theoretical prediction is given by eqn.~(\ref{MedGauss}) together with eqn.~(\ref{eq:SA_gaussian}).
Panels (a) and (b) use the initial state $\ket{\psi}=\ket{0}$, which is a system eigenstate with energy $E_0=-5$. Panels (c) and (d) consider the superposition $\ket{\psi}=\sfrac{1}{2}\ket{1}+\sfrac{\sqrt{3}}{2}\ket{5}$, which is a linear combination of the eigenstate $\ket{1}$ with eigenvalue $E_1=-1$, and the eigenstate $\ket{5}$ with eigenvalue $E_5=3$. The results empirically demonstrate that the local maxima of the SA correspond to the probability of measuring the input state at the peak energy. In panels (a) and (c), the controlled time evolution is implemented with an ancilla qubit ($d=2$), whereas in panels (b) and (d) an ancilla qutrit ($d=3$) is employed.

A visual inspection of the figure indicates that, when comparing the ancilla qutrit with the ancilla qubit, the spectral peak becomes narrower and the fluctuations are reduced for energy values away from the peak position. To quantify these effects, we introduce two metrics. First, we consider the relative difference between the SA obtained using an ancilla qudit and that obtained with an ancilla qubit, which characterizes the narrowing of the peak. Second, in the region where the SA is approximately zero, the fluctuation is quantified by the standard deviation of the measured points. 

\subsubsection*{Peak Narrowing}
\begin{figure}[!ht]
\begin{tabular}{c c c c}
\includegraphics[scale=0.24] {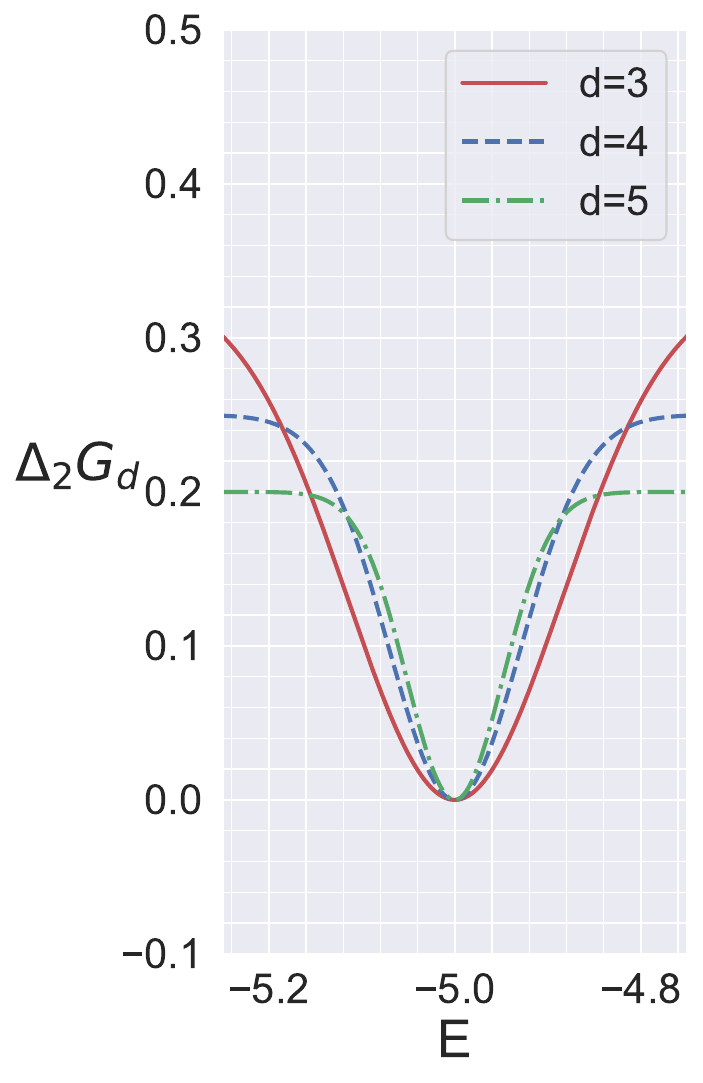} & 
\includegraphics[scale=0.24]{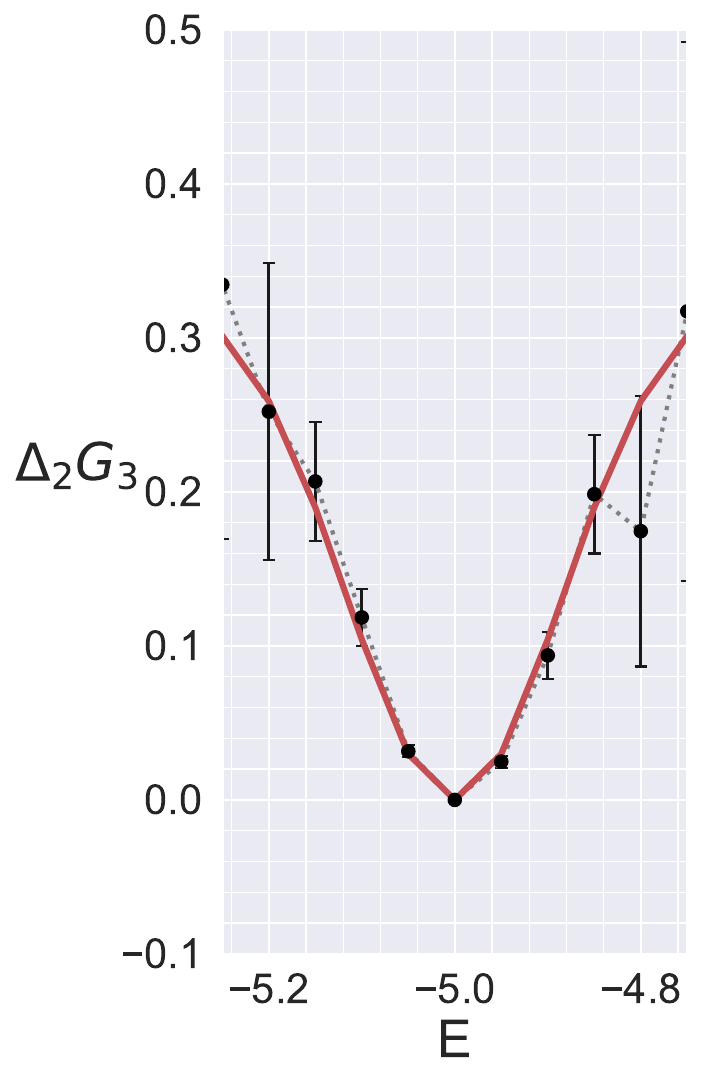}  & 
\includegraphics[scale=0.24] {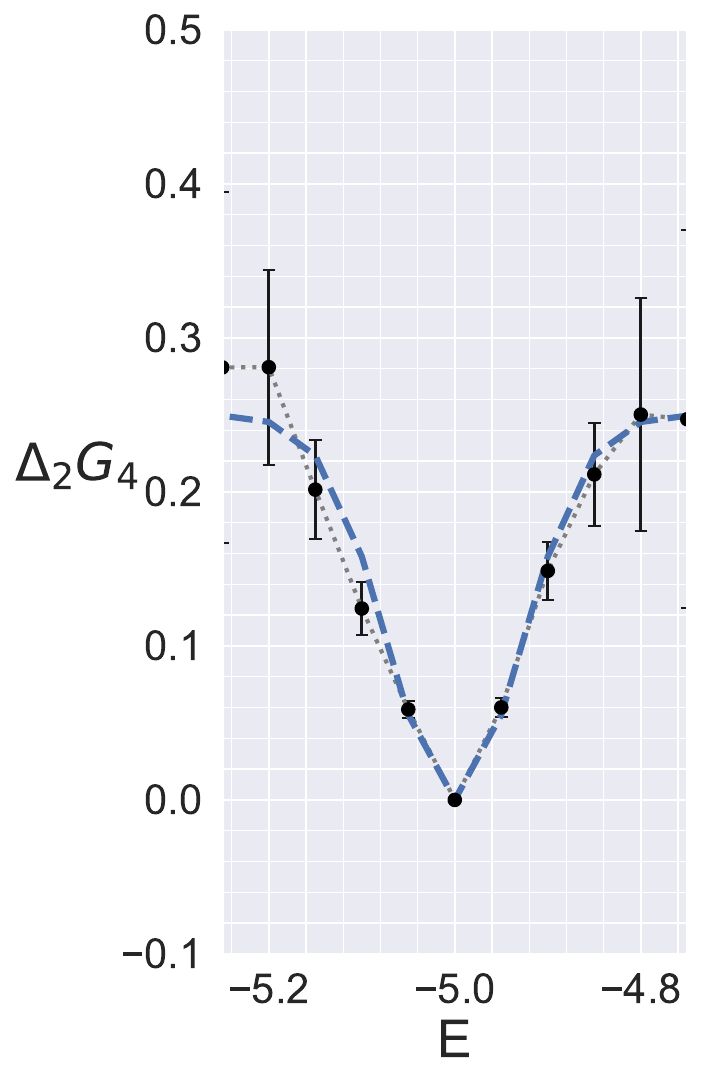} &
\includegraphics[scale=0.24]{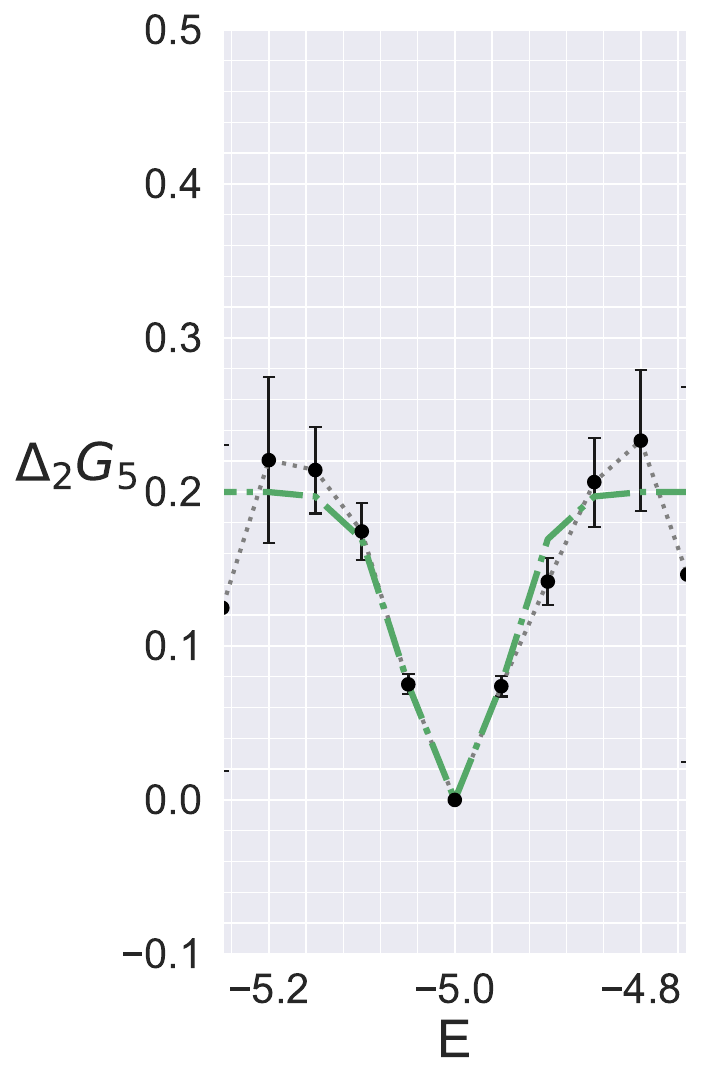} \\ 
 (a) &  (b) & (c) & (d)
\end{tabular} 
\caption{Relative difference of the spectral amplitude obtained with a $d$-dimensional ancilla qudit with respect to the ancilla qubit case ($d=2$) for the one-dimensional spin-$\tfrac{1}{2}$ Ising model with input state $\ket{\psi}=\ket{0}$. The Gaussian distribution used in the averaging has $\sigma = 5$. Panel (a) shows the theoretical prediction for different ancilla dimensions (see eqn.~(\ref{eq:relDiff_x})). Panels (b)--(d) display the corresponding numerical results for a periodic chain of five spins, obtained from $500$ independent realizations of the evolution time for each energy value: (b) $d=3$, (c) $d=4$, and (d) $d=5$.}
\label{fig:IsingHalf_diff}
\end{figure}

We first analyze the peak narrowing. For an eigenstate input, $\ket{\psi}=\ket{x}$, eqn.~(\ref{eq:relDiff}), which governs the relative difference, can be written as
\begin{equation}
\Delta_2 G_d(E,x) =
\frac{1 - e^{-\frac{\sigma^2 (d^2 - 2d)(E_x - E)^2}{2}}}{d}.
\label{eq:relDiff_x}
\end{equation}
This relation is constrained to the interval $0 \le \Delta_2 G_d(E,x) \le d^{-1}$, which implies $G_2(E,x) \ge G_d(E,x)$. This behavior is illustrated in Fig.~\ref{fig:IsingHalf_diff} for $\ket{x}=\ket{0}$. Panel (a) shows a comparison for $d=3,\ 4,$ and $5$ via a directly plot of eqn.~(\ref{eq:relDiff_x}). As discussed previously, the perturbation in the peak decay tends to zero as $d$ increases. Panels (b), (c), and (d) demonstrate that the numerical results corroborate the theoretical prediction, for $d=3,\ 4, \text{ and } 5$, respectively. Due to error propagation, which leads to error bars that are inversely proportional to $G_2(E,x)$, we display only the peak-decay region. The implementation employing an ancilla qutrit shows the strongest peak narrowing.

\subsubsection*{Fluctuation}
We quantify the fluctuations by the standard deviation of the numerical data across different energy values. Since all energy values in the region far from the eigenvalue obey the same statistics, we expect this dispersion to be comparable to the error bars of the individual points, which correspond to the standard error of the ensemble average. Because $h_d(E,\psi|t)$ is, in general, complex, the ensemble averages of its real and imaginary parts were computed separately. For an eigenstate input, $\ket{\psi}=\ket{x}$, the average of the imaginary part vanishes, besides that Appendix~\ref{sec:variance} also shows that, for $d>2$, the maximum standard deviations of both parts are
\begin{equation}
   \mathrm{std}_m[\mathrm{Re}(h_d(E,x))] = \mathrm{std}_m[\mathrm{Im}(h_d(E,x))]
   = \frac{1}{d}\sqrt{\frac{d^2 - 2d + 2}{2N_t}} .
   \label{eq:stdReal_h}
\end{equation}
For the qubit implementation, $d=2$, the two frequency components of the Rodeo kernel coincide, the signal is purely real, and the same appendix yields
\begin{equation}
   \mathrm{std}_m[h_2(E,x)] = \sqrt{\frac{1}{2N_t}} .
   \label{eq:stdReal_h2}
\end{equation}
Comparison of eqns.~(\ref{eq:stdReal_h}) and (\ref{eq:stdReal_h2}) shows that, for any finite $d>2$, the qudit implementation exhibits smaller fluctuations than the qubit implementation, with both coinciding in the limit $d\to\infty$. Since $(d^2-2d+2)/d^2$ increases monotonically for $d\ge 3$, the minimum is attained for the ancilla qutrit, which yields a theoretical reduction of approximately $25\%$ in the standard deviation relative to the qubit. This reduction originates from the splitting of the signal into two components of distinct frequencies, $\omega_x$ and $(d-1)\omega_x$, whose cross-correlation vanishes upon time averaging; it is therefore an intrinsic feature of ancillary of dimension $d>2$.

In Table~\ref{tab:error}, we compare the theoretical predictions with the numerical estimates. The first column indicates the ancilla qudit dimensionality. The second column corresponds to the direct evaluation of eqn.~(\ref{eq:stdReal_h}), considering $N_t = 500$, which is the number of realizations used in our numerical simulations. The third column reports the average of the error bars in the region $G_d(E,x) < 0.1$ (the uncertainties were computed from five independent simulations). We observe good agreement between the theoretical and numerical values. The fourth column shows the fluctuation metric, previously discussed. Finally, the last column presents the reduction ratio, defined as the relative difference between the numerical estimates of the fluctuations for the ancilla-qudit and ancilla-qubit implementations. 

\begin{table}[h]
    \caption{Comparison of theoretical and numerical fluctuation estimates for different ancilla dimensions. The columns report the ancilla dimension \(d\), the theoretical standard deviation evaluated from eqn.~(\ref{eq:stdReal_h}) for \(N_t=500\) realizations, the mean numerical error-bar magnitude averaged over the region \(G_d(E,x)<0.1\) across five independent simulations, the fluctuation metric defined in the text, and the relative fluctuation reduction with respect to the qubit-ancilla implementation.}
    \label{tab:error}
    \begin{tabular}{@{}ccccc@{}}
    \toprule
     \hspace{0.2cm}   $d$  \hspace{0.2cm}  &\hspace{0.2cm}  $\mathrm{std}_{m}$  \hspace{0.2cm} &  \hspace{0.2cm}Error bar   \hspace{0.2cm} &  Fluctuation \hspace{0.2cm}&\hspace{0.2cm} reduction ratio \hspace{0.2cm}  \\ 
     \hline
        $2$     & $\approx 0.03162$ & $0.03164(1)$  &   $0.0338(2)$ &  -- \\
        $3$  & $\approx 0.02357$ & $0.02354(1)$  &  $0.0276(3)$ & $18.3\%$ \\
        $4$    & $0.025$  & $0.02497(1)$      &   $0.0278(5)$ &  $17.8\%$\\
        $5$    &  $\approx 0.02608$&$0.02600(1)$    &    $0.0292(1)$ &  $13.6\%$\\
    $\infty$   & $\approx 0.03162$  &     --      & -- & -- \\
    \botrule
    \end{tabular}
\end{table}

Although discrepancies between the theoretical prediction and the numerical estimates, ranging from $7\%$ to $17\%$, are observed, we interpret the fluctuations of the data points across different energies as reflecting the standard deviation indicated by the error bars. As conjectured, 
the ancilla qutrit implementation provides some noise reduction and improved spectral resolution, yielding an $18\%$ reduction in fluctuations. Although smaller than the theoretically predicted $25\%$, this reduction remains significant.

\subsection{1D Spin-$1$ Ising Model}
\label{sec.spin1}

As an illustrative example of a multi-level quantum system, we consider the one-dimensional spin-$1$ Lenz–Ising model, for which the local Hilbert-space dimension is $d'=3$. In this case, the spin operator along the  $z$-direction can be written as
\begin{equation}
    S_3^z =
    \begin{pmatrix}
        1 & 0 & 0 \\
        0 & 0 & 0 \\
        0 & 0 & -1
    \end{pmatrix}
    = \ketbra{0}{0} - \ketbra{2}{2}.
\end{equation}
In Fig.~\ref{fig:IsingOne}~(a), we present the SA for a chain of three particles with the input state $\ket{15}$. The state labels are written in decimal (radix-10) notation, but they can be interpreted in terms of spin configurations by converting the index to balanced ternary notation. Because the particle positions are labeled from left to right, the ternary digits must be read in reverse order. Under this convention, the state $\ket{15}$ corresponds to the configuration $\ket{0,-1,1}$, or in a pictorial $z$-component representation,  $\ket{\circ,\downarrow,\uparrow}$. The corresponding energy eigenvalue is $E_{15}=+1$, which the SA accurately recovers. The results presented here were obtained by sampling $500$ realizations of the evolution time from a normal distribution with standard deviation $\sigma = 10$. 

\subsubsection*{Microcanonical Protocol}

\begin{figure}[!ht]
\begin{tabular}{c c}
\includegraphics[scale=0.22]{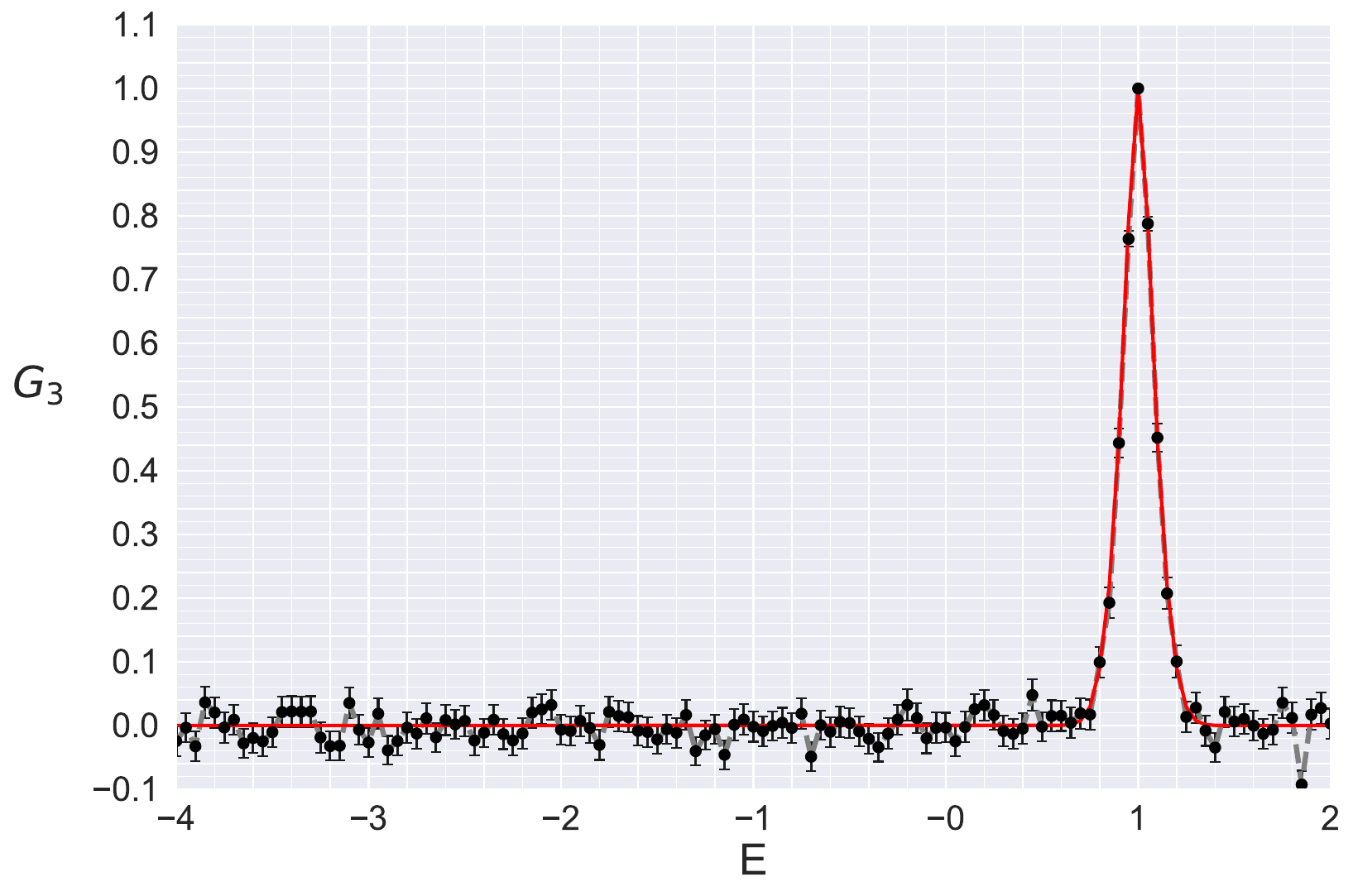}  & 
\includegraphics[scale=0.23]{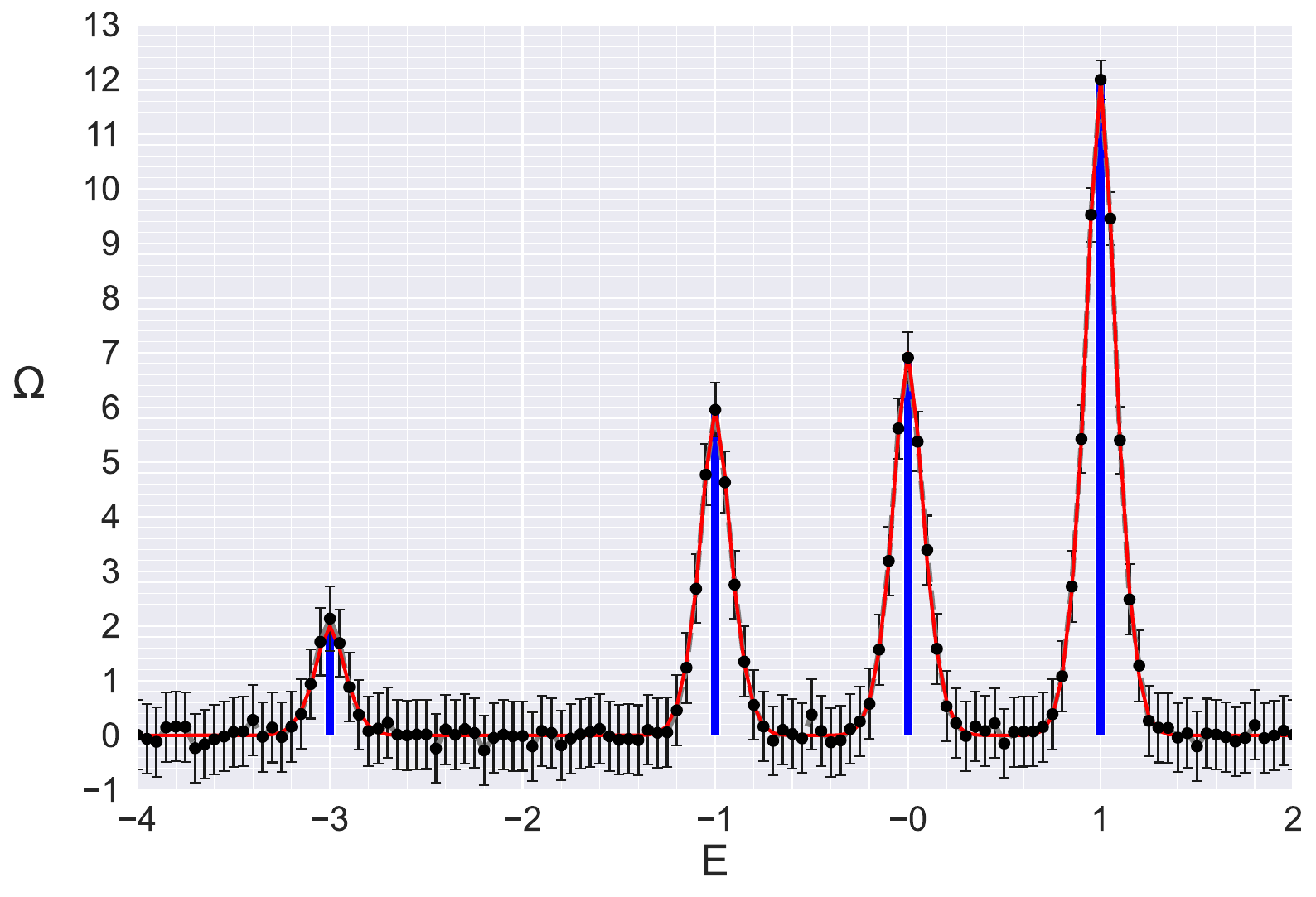} \\
 (a) &  (b)
\end{tabular} 
\caption{Numerical evaluation of the Rodeo algorithm applied to the one-dimensional spin-$1$ Ising model with $N=3$ spins. The evolution times are sampled from a Gaussian distribution with standard deviation $\sigma = 10$. (a) Spectral amplitude for the computational-basis input state $\ket{\psi} = \ket{15}$, averaged over $500$ independent evolution-time samples for each energy value. (b) Number of states, obtained by summing the spectral amplitude over all computational-basis input states, following the protocol described in Ref.~\cite{Rocha024}. }
\label{fig:IsingOne}
\end{figure}

Fig.~\ref{fig:IsingOne}~(b) shows the resulting number of states (NoS), obtained using the protocol described in Ref.~\cite{Rocha024}. As discussed in Sec.~\ref{sec:entropic}, the NoS is estimated by summing the spectral amplitudes over all computational-basis states used to represent the system. The agreement between the vertical blue line and the spectral peak positions demonstrates that the protocol correctly reproduces the degeneracies of the energy levels. This result also illustrates the accumulation of statistical fluctuations at energies far from the eigenvalues, where noise contributions from the individual input states add up. In contrast, using a single homogeneous superposition of computational-basis states avoids this accumulation of noise, as shown below.

\subsubsection*{Single-State Microcanonical Protocol}
\begin{figure}[!ht]
\begin{tabular}{c c}
\includegraphics[scale=0.22] {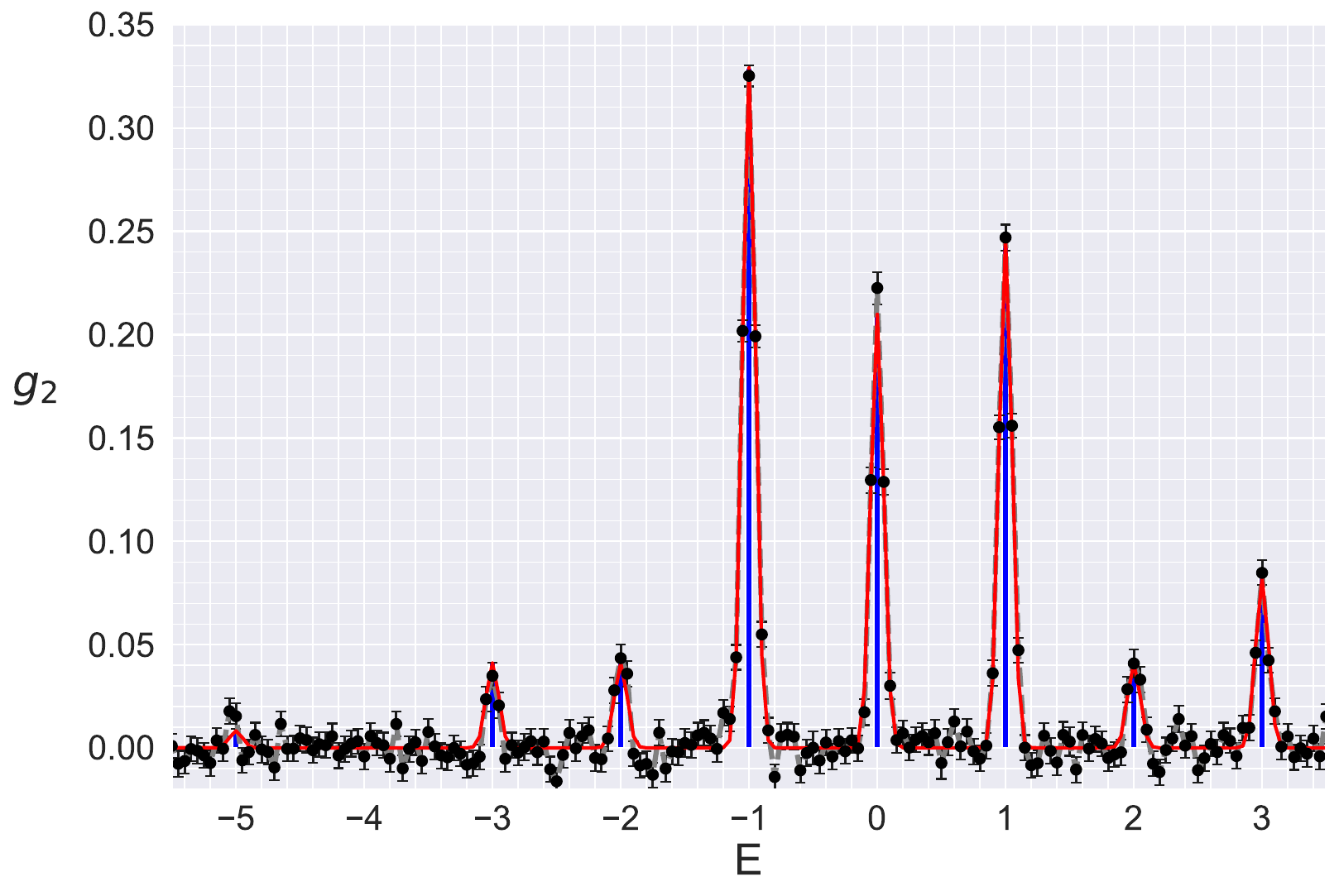} & 
\includegraphics[scale=0.22]{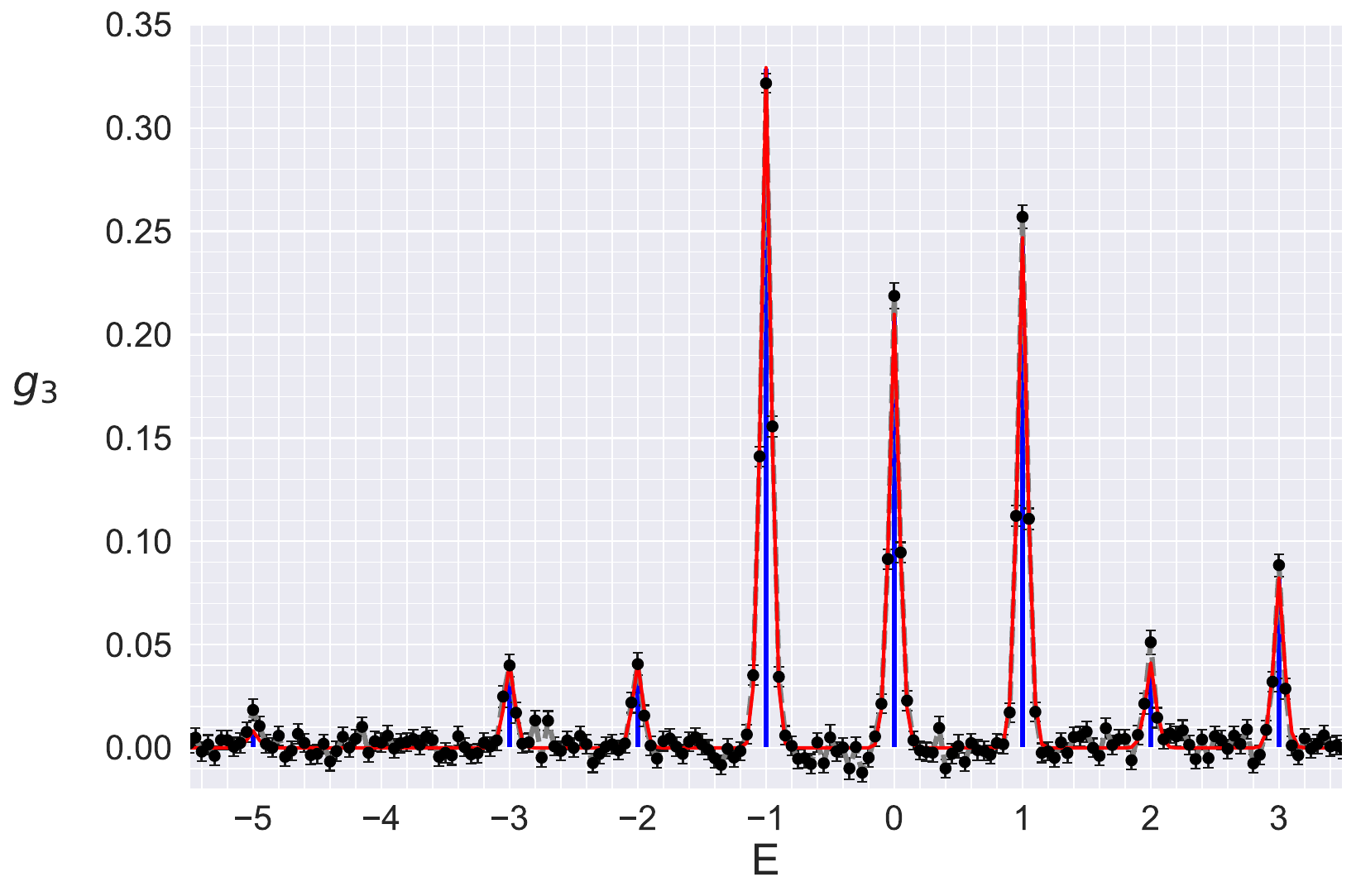} \\ 
 (a) &  (b)
\end{tabular} 
\caption{ Single-state microcanonical protocol for the Rodeo Algorithm applied to the one-dimensional spin-$1$ Ising model. We consider a periodic chain of five spins. The black dots represent the average from $3000$ independent realizations of the evolution time for each energy value, while the error bars denote the corresponding standard deviations. The evolution times are sampled from a Gaussian distribution with $\sigma = 20$. In panel (a) the controlled time evolution is implemented with an ancilla qubit ($d=2$), while in panel (b) an ancilla qutrit ($d=3$) is employed.}
\label{fig:DOS}
\end{figure}

Fig.~\ref{fig:DOS} shows the result for the single-state microcanonical protocol of the Rodeo algorithm for the one-dimensional spin-$1$ Ising model with $N=5$ particles. Panels (a) and (b) correspond to the ancilla qubit and ancilla qutrit implementations, respectively. As expected, the qutrit implementation achieves some noise reduction and improved spectral resolution. The SA in this protocol is interpreted as the DoS, here computed from $3000$ independent realizations of the evolution time for each energy value, sampled from a Gaussian distribution with $\sigma = 20$. The results clearly exhibit the Gaussian smoothing of the energy spectrum, confirming that the protocol effectively performs a Gaussian convolution of the spectral density. 

This figure illustrates the compromise between computational cost and statistical precision. Although the qutrit implementation provides some degree of noise reduction, identifying the ground-state degeneracy remains challenging for $N_t = 3000$, even when using an ancilla qutrit. Despite avoiding the accumulation of noise from multiple input states, the single-state protocol yields a signal normalized by the total Hilbert-space dimension, i.e., \(g_d(E)=\Omega_\sigma(E)/d^{\prime N}\). Consequently, the signal-to-noise ratio may decrease exponentially, potentially requiring exponentially many time samples \(N_t\) to resolve low-entropy states, even with qudit-induced variance reduction. Nevertheless, partial knowledge of the DoS is often sufficient to describe the thermodynamic behavior of a physical system, particularly when the analysis focuses on a narrow region of the thermodynamic coordinate space. 


\section{Conclusion}
\label{sec:conclusion}

In this work, we developed a qudit formulation of the Rodeo algorithm with a general $d$-level ancilla and investigated its theoretical and numerical properties by deriving the full evolution of the joint ancilla--system state. In the context of state preparation, we showed that the success-probability distribution becomes progressively narrower as $d$ increases. We further established that its width can be matched to that of the qubit implementation by rescaling the standard deviation of the evolution-time distribution according to $\sigma_d=\sigma_2\sqrt{3/(d^2-1)}$. We also demonstrated that outcomes with $n\neq 0$ are informative, as their maxima occur at $\omega_x t=-2\pi n/d \pmod{2\pi}$. These outcomes therefore encode both the magnitude and sign of the detuning and could be exploited to adaptively update the trial energy.

At the level of expectation values, measuring the clock operator $Z_d$ yields the Rodeo kernel, which acts as a two-frequency interferometer: its dominant contribution reproduces the Loschmidt amplitude, while a subleading term introduces harmonic contamination. This structure arises from the periodic boundary conditions of the finite-dimensional ancilla basis. It appears only for $d>2$ and vanishes in the limit $d\to\infty$, ensuring that the qudit-induced corrections to the spectral amplitude remain perturbative and are most pronounced for the qutrit. Under Gaussian time sampling, these corrections sharpen the spectral peak and contribute to noise reduction. Relative to the qubit-ancilla implementation, the predicted noise reduction for the qutrit is $25\%$. For the spin-$\tfrac{1}{2}$ Ising chain, our simulations yield a reduction of approximately $18\%$, with both the qubit and qutrit results in quantitative agreement with the analytical predictions.

We also introduced a single-state microcanonical protocol in which a homogeneous superposition of energy eigenstates serves as the input state. The resulting spectral amplitude corresponds to a Gaussian convolution of the density of states (DoS), allowing the microcanonical entropy to be estimated through a single energy sweep without explicitly enumerating the exponentially large computational basis. For the spin-1 Ising chain, the protocol successfully reproduces the energy-level degeneracies, while the qutrit ancilla again provides reduced noise. However, because the signal scales as $d'^{-N}$, resolving low-entropy regions, particularly those associated with ground-state degeneracies, remains challenging. Even with the variance reduction afforded by qudits, the required number of time samples may grow exponentially with system size. Moreover, preparing a homogeneous superposition of energy eigenstates may be impractical for general Hamiltonians. Thus, direct implementation of this protocol is restricted to Hamiltonians whose energy eigenbasis coincides with the computational basis, as in the Ising models considered here. Motivated by this construction, we proposed an alternative scheme for approximating the DoS that extends to general Hamiltonians~\cite{Rocha2026}.

Overall, our results establish multi-level qudits as a promising framework for quantum spectral analysis and thermodynamic characterization. By exploiting their additional degrees of freedom, the qudit-based Rodeo algorithm offers potential advantages in spectral resolution and noise suppression, while providing a natural approach to investigating quantum systems with multi-level degrees of freedom.

\appendix
\section{Projection Amplitude}
\label{amplitude}

To prove eqn.~(\ref{eq:dirchlet_kernel}), we define
\begin{equation}
\theta_x(n)
= \omega_x t + \frac{2\pi n}{d}.
\label{eq:def_theta}
\end{equation}

Using the finite geometric series identity
\begin{equation}
\sum_{\ell=0}^{d-1} e^{-i \ell \theta}
=
\frac{1 - e^{-i \theta d}}{1 - e^{-i \theta}}
=
e^{-i \frac{(d-1)\theta}{2}}
\frac{\sin \left(\theta d / 2\right)}
{\sin \left(\theta / 2\right)},
\label{Dirichletkernel}
\end{equation}
where $\sin\theta = (e^{i\theta}-e^{-i\theta})/(2i)$, the above expression can be recognized as the Dirichlet kernel~\cite{He2019} for non-negative integer $\ell$.

The projection amplitude, eqn.~(\ref{eq:projAmpl}), can then be written as
\begin{equation}
A(n)
=
\frac{1}{d}
\sum_x
|c_x|^2
e^{-i \frac{(d-1)\theta_x(n)}{2}}
\frac{\sin \left(\theta_x(n) d / 2\right)}
{\sin \left(\theta_x(n) / 2\right)}.
\end{equation}
Substituting the definition of $\theta_x(n)$, eqn.~(\ref{eq:def_theta}), into the numerator gives
\begin{equation}
\sin \left(\omega_x \frac{t d}{2} + \pi n\right)
=
e^{i \pi n}
\sin \left(\omega_x \frac{t d}{2}\right),
\end{equation}
which then leads to eqn.~(\ref{eq:dirchlet_kernel}):
\begin{equation}
A(n)
=
\frac{e^{i \frac{\pi n}{d}}}{d}
\sum_x
|c_x|^2
e^{-i \omega_x' \frac{t}{2}}
\frac{\sin \left(\omega_x \frac{t d}{2}\right)}
{\sin \left(\omega_x \frac{t}{2} + \frac{\pi n}{d}\right)},
\end{equation}
where $\omega_x' = (d-1)\omega_x$.


\section{Ancilla Level Measurement Probability Behavior}
\label{sec:success}
In this appendix, we analyze the behavior of the ancilla outcome distribution given in eqn.~(\ref{eq:p_dn_x}), reproduced here for convenience:
\[
P_d(n) \approx \frac{1}{d^2}
\left|
\frac{
\sin \left(\omega_x \frac{td}{2}\right)
}{
\sin \left(\omega_x \frac{t}{2} + \frac{\pi n}{d}\right)
}
\right|^2.
\]
\subsection{Success Probability Width Control}
In this section, we derive the condition under which the qudit implementation matches the qubit implementation's sensitivity to deviations from resonance. 
For small detuning, denoted $\delta \omega_x$, eqn.~(\ref{eq:p_dn_x}) can be expanded in a Taylor series and approximated by retaining the leading two terms:
\begin{equation}
P_d(0) \approx 1 - \frac{(d^2 - 1)t^2}{12}(\delta\omega_x)^2.
\end{equation}
Averaging over the Gaussian distribution of evolution times (similar to eqn.~(\ref{eq:SA_definition})), with zero mean and standard deviation \(\sigma\), yields
\begin{equation}
\overline{P_d(0)} \approx
1 - \frac{(d^2 - 1)\sigma^2}{12}(\delta\omega_x)^2.
\end{equation}
To quantify the allowable deviation in the success probability, let us introduce a tolerance \(\varepsilon\), such that the deviation from unity satisfies
\begin{equation}
\varepsilon = \frac{(d^2 - 1)\sigma^2}{12}
(\delta\omega_x)^2.
\end{equation}
Solving for the corresponding detuning scale gives
\begin{equation}
\delta\omega_x =
\frac{1}{\sigma}
\sqrt{\frac{12\varepsilon}{d^2 - 1}}.
\end{equation}

For a qubit ancilla (\(d=2\)), this expression reduces to
\begin{equation}
\delta\omega_x = \frac{\sqrt{4\varepsilon}}{\sigma}.
\end{equation}

Consequently, the standard deviation for an ancilla qudit of dimension \(d\) can be chosen to reproduce the same detuning scale as in the qubit implementation. Equating the two expressions for \(\delta\omega_x\), we obtain
\begin{equation}
    \frac{\sqrt{4\varepsilon}}{\sigma_2}
=
\frac{1}{\sigma_d}
\sqrt{\frac{12\varepsilon}{d^2-1}},
\end{equation}
which leads to eqn.~(\ref{eq:ratio_sigma}),
\[
\sigma_d = \sigma_2
\sqrt{\frac{3}{d^2-1}}.
\]
Here, \(\sigma_d\) denotes the standard deviation of the evolution-time distribution for an implementation using an ancilla qudit of dimension \(d\), whereas \(\sigma_2\) denotes the corresponding standard deviation for a qubit ancilla. This relation provides a simple analytical criterion for adjusting the distribution width when generalizing the algorithm to higher-dimensional ancillas, within the small-detuning approximation.

\subsection{Extrema of the Outcome Distribution $P_d(n)$}
\label{app:extrema}

In this section we locate the global maxima of eqn.~\eqref{eq:p_dn_x}.
\subsubsection*{Reduction to the Fej\'er kernel}

We introduce the shifted phase
\begin{equation}
\alpha_n = \frac{\omega_x t}{2} + \frac{\pi n}{d}.
\label{eq:alpha_n}
\end{equation}
Since $\omega_x t d/2 = d\alpha_n - \pi n$, the numerator satisfies
$\sin(\omega_x t d/2) = (-1)^n \sin(d\alpha_n)$, and the sign disappears
under the modulus. Therefore
\begin{equation}
P_d(n) = F_d(\alpha_n),
\qquad
F_d(\alpha) \equiv \frac{1}{d^2}\,\frac{\sin^2(d\alpha)}{\sin^2\alpha}
= \left| \frac{1}{d}\sum_{\ell=0}^{d-1} e^{2i\ell\alpha} \right|^2 ,
\label{eq:fejer}
\end{equation}
where the last equality follows from the finite geometric series used in
Appendix~A. The function $F_d$ is the Fej\'er kernel, i.e., the squared
modulus of the normalized Dirichlet kernel. All outcomes $n$ are thus
described by the same function, shifted by $\pi n/d$; the distribution is
$\pi$-periodic in $\alpha_n$ and normalized,
$\sum_{n=0}^{d-1} P_d(n) = 1$.

\subsubsection*{Principal Maxima}

By the triangle inequality applied to eqn.~\eqref{eq:fejer},
\begin{equation}
F_d(\alpha) \le \left( \frac{1}{d}\sum_{\ell=0}^{d-1} 1 \right)^2 = 1 ,
\end{equation}
with equality if and only if all phases $e^{2i\ell\alpha}$ coincide, that
is, $e^{2i\alpha}=1$, or $\alpha = k\pi$ with $k\in\mathbb{Z}$. At these
points the ratio $\sin(d\alpha)/\sin\alpha \to \pm d$, and $F_d = 1$.
Using eqn.~\eqref{eq:alpha_n}, the principal maxima of the $n$-th outcome are
\begin{equation}
\omega_x t = 2\pi\left(j - \frac{n}{d}\right),
\qquad
P_d(n)\big|_{\max} = 1 ,
\qquad j \in \mathbb{Z},
\label{eq:pdn_max}
\end{equation}
that is, $\omega_x t = -2\pi n/d \pmod{2\pi}$, demonstrating eqn.~(\ref{eq:max_pdn}). In terms of the trial
energy, with $\omega_x = E_x - E$,
\begin{equation}
E^{(n)}_{\max} = E_x + \frac{2\pi}{t}\left(\frac{n}{d} - j\right).
\label{eq:Emax_n}
\end{equation}
For $n = 0$ and $j=0$ one recovers the resonance
condition $E = E_x$. Because of normalization, whenever one outcome reaches
$P_d(n)=1$, all others vanish: the principal maximum of each outcome
coincides with zeros of the remaining $d-1$ outcomes.

\section{Calculation of Variance and Standard Deviation}
\label{sec:variance}
In this section, we analyze the variance of the spectral amplitude. By definition, the variance is
\begin{equation}
\mathrm{Var}[h(E,\psi|t)]
\equiv
\overline{h^2(E,\psi|t)}-
\Big[\overline{h(E,\psi|t)}\Big]^2 ,
\label{eq:var_def}
\end{equation}
where the overline denotes averaging with respect to the Gaussian
time-sampling distribution introduced in
Sec.~\ref{sec:Ensemble}. 
\subsection{Real Part}
\label{sec:varReal}
The real part of eqn.~(\ref{H_echo}) satisfies
\begin{equation}
    \mathrm{Re}(K_d(\omega_x,t)) =
    \frac{d-1}{d} \cos{(\omega_x t)} +
    \frac{1}{d} \cos{(\omega_x' t)}.
    \label{eq:realSA}
\end{equation}
Using
\begin{equation}
    \frac{1}{\sigma \sqrt{2\pi}}
    \int^{\infty}_{-\infty}
    \cos(\omega t)\,
    e^{-\frac{(t-\mu)^2}{2\sigma^2}} \mathrm{d}t
    =
    e^{-\frac{\sigma^2\omega^2}{2}}\cos(\omega\mu),
\end{equation}
the Gaussian average of eqn.~(\ref{eq:realSA}) for $\mu=0$ reduces to
eqn.~(\ref{eq:SA_gaussian}). Consequently,
squaring this equation  yields
\begin{equation}
    \Big[\overline{\mathrm{Re}(h_d(E,\psi|t)})\Big]^2 = G_d^2(E,\psi) =
\sum_{x,y} |c_x|^2 |c_y|^2
\big[\mathcal{T}_P(\omega_x)\big]\big[\mathcal{T}_P(\omega_y)\big],
\end{equation}
where
\begin{align} \nonumber
 \big[\mathcal{T}_P(\omega_x)\big]\big[\mathcal{T}_P(\omega_y)\big] = 
 e^{-\frac{\sigma^2 (\omega_x^2 +\omega_y^2)}{2}}
  &\left[ 1 - \left( \frac{1 - e^{-\frac{\sigma^2 (\omega_x'^2 - \omega_x^2)}{2}} }{d} \right) \right] \times \\ 
  & \left[ 1 - \left( \frac{1 - e^{-\frac{\sigma^2 (\omega_y'^2 - \omega_y^2)}{2}} }{d} \right) \right].
\label{eq:squareSA}
\end{align}

For $\ket{\psi}=\ket{x}$ this relation becomes
\begin{equation}
\Big[\overline{\mathrm{Re}(h_d(E,x|t))}\Big]^2 = \big[\mathcal{T}_P(\omega_x)\big]^2  =
 e^{-\sigma^2 \omega_x^2}
  \left[
  1 -
  \left(
  \frac{1 - e^{-\frac{\sigma^2 (\omega_x'^2 - \omega_x^2)}{2}}}{d}
  \right)
  \right]^2 .
  \label{eq:sectermvar}
\end{equation}

On the other hand, squaring eqn.~(\ref{eq:realSA}) gives
\begin{align} \nonumber
    \mathrm{Re}(K_d(\omega_x,t)) \mathrm{Re}(K_d(\omega_y,&t))
    =
    \frac{1}{d^2}\Bigg\{
    (d-1)^2 \cos{(\omega_x t)}\cos{(\omega_y t)}
    + \cos{(\omega_x' t)}\cos{(\omega_y' t)} \\
    &+ (d-1)\Big[
    \cos{(\omega_x t)}\cos{(\omega_y' t)}
    + \cos{(\omega_y t)}\cos{(\omega_x' t)}
    \Big]\Bigg\}.
\end{align}

Using the identity
$\cos a \cos b = [\cos(a-b)+\cos(a+b)]/2$
and considering $\ket{\psi}=\ket{x}$, we obtain
\begin{align} \nonumber
    \big[\mathrm{Re}(K_d(\omega_x,t))\big]^2
    =
    \frac{1}{d^2}\Big[&
    (d-1)^2 \frac{1+\cos(2\omega_x t)}{2}
    +
    \frac{1+\cos(2\omega_x' t)}{2} \\
    &+
    (d-1)\big(
    \cos[(\omega_x'-\omega_x)t]
    +
    \cos[(\omega_x'+\omega_x)t]
    \big)
    \Big].
\end{align}

Performing the Gaussian average with $\mu=0$ gives
\begin{align}\nonumber
     \overline{[\mathrm{Re}(K_d(\omega_x,t))]^2}
     =
     \frac{1}{d^2}\Big[&
     (d-1)^2 \frac{1+e^{-2\sigma^2\omega_x^2}}{2}
     +
     \frac{1+e^{-2\sigma^2\omega_x'^2}}{2} \\
     &+
     (d-1)\left(
     e^{-\frac{\sigma^2}{2}(\omega_x'-\omega_x)^2}
     +
     e^{-\frac{\sigma^2}{2}(\omega_x'+\omega_x)^2}
     \right)
     \Big].
\label{eq:varRealSAgeral}
\end{align}
For $\ket{\psi}=\ket{x} \Longrightarrow \overline{[\mathrm{Re}(h_d(E,x|t))]} = \overline{[\mathrm{Re}(K_d(\omega_x,t))]^2}$. Therefore, subtracting eqn.~(\ref{eq:sectermvar}) from the relation above leads to
\begin{align} \nonumber
   \mathrm{Var}[\mathrm{Re}(h_d(E,x|t))]  =
   \frac{1}{d^2}\Big[&
     (d-1)^2 \frac{1+e^{-2\sigma^2\omega_x^2}}{2}
     +
     \frac{1+e^{-2\sigma^2\omega_x'^2}}{2} \\ \nonumber
     &+
     (d-1)\left(
     e^{-\frac{\sigma^2}{2}(\omega_x'-\omega_x)^2}
     +
     e^{-\frac{\sigma^2}{2}(\omega_x'+\omega_x)^2}
     \right)
     \Big] \\
   &- e^{-\sigma^2 \omega_x^2}
  \left[
  1 -
  \left(
  \frac{1 - e^{-\frac{\sigma^2 (\omega_x'^2 - \omega_x^2)}{2}}}{d}
  \right)
  \right]^2
    \label{eq:varqudit}
\end{align}
For $E=E_x$, i.e., $\omega_x=\omega_x' = 0$, it leads to $\mathrm{Var}[\mathrm{Re}(h_d(E_x,x))]  = 0$. This demonstrates that the error bar at the peak position is equal to zero.

For the qubit implementation, \(\omega'=\omega_x\), eqn.~(\ref{eq:varqudit}) simplifies to
\begin{equation}
   \mathrm{Var}[\mathrm{Re}(h_2(E,x|t))]
   = \frac{1}{2}\left(1-e^{-\sigma^2\omega_x^2}\right)^2.
   \label{eq:varqubit}
\end{equation}

We can identify the maximal variance in the region far from an eigenvalue, where \(|\omega_x|\) is sufficiently large that the exponential terms vanish. The variance approaches
\begin{equation}
   \mathrm{Var}_m[\mathrm{Re}(h_2(E,x|t))]=\frac{1}{2}
\end{equation}
for the qubit implementation (eqn.~(\ref{eq:varqubit})), whereas for \(d>2\), eqn.~(\ref{eq:varqudit}) gives
\begin{equation}
   \mathrm{Var}_m[\mathrm{Re}(h_d(E,x|t))]
   =\frac{(d-1)^2+1}{2d^2}.
\end{equation}
This expression is minimized at \(d=2\). Since it holds only for \(d>2\), the minimum occurs at \(d=3\). Moreover,
\[
   \lim_{d\to\infty}\mathrm{Var}_m[\mathrm{Re}(h_d(E,x|t))]
   =\frac{1}{2},
\]
recovering the asymptotic variance of the qubit implementation.

The maximal standard deviation for the qubit is
\begin{align}
   \mathrm{std}_m[\mathrm{Re}(h_2(E,x|t))]
   = \sqrt{\frac{1}{2N_t}},
\end{align}
and for $d>2$ is then:
\begin{align}
   \mathrm{std}_m[\mathrm{Re}(h_d(E,x|t))]
   =
   \frac{1}{d}\sqrt{\frac{d^2 - 2d +2}{2N_t}},
\end{align}
where $N_t$ is the number of realizations of the evolution time. The vanishing standard deviation at the peak position suggests the potential for an adaptive sampling strategy, in which the number of algorithmic cycles is reduced in regions where the variance approaches its maximum. Since no discernible spectral features are present in these regions, computational resources can be allocated more efficiently by prioritizing regions of greater spectral interest.

\subsection{Imaginary Part}
\label{sec:varImag}

The imaginary part of eqn.~(\ref{H_echo}) satisfies
\begin{equation}
    \mathrm{Im}(K_d(\omega_x,t)) =
    -\frac{d-1}{d} \sin{(\omega_x t)} +
    \frac{1}{d} \sin{(\omega_x' t)}.
    \label{eq:imgSA}
\end{equation}
For $\mu=0$, the Gaussian average vanishes because the integrand is an odd function, i.e.,
\begin{equation}
    \frac{1}{\sigma \sqrt{2\pi}}
    \int^{\infty}_{-\infty}
    \sin(\omega t)
    e^{-\frac{t^2}{2\sigma^2}} \mathrm{d}t
    = 0.
\end{equation}
Squaring eqn.~(\ref{eq:imgSA}) yields
\begin{align} \nonumber
    \mathrm{Im}(K_d(\omega_x,t)) \mathrm{Im}(K_d(\omega_y,&t))
    = 
    \frac{1}{d^2} \Bigg\{
    (d-1)^2 \sin{(\omega_x t)}\sin{(\omega_y t)}
    + \sin{(\omega_x' t)}\sin{(\omega_y' t)} \\
    &- (d-1)\Big[
    \sin{(\omega_x t)}\sin{(\omega_y' t)}
    +
    \sin{(\omega_y t)}\sin{(\omega_x' t)}
    \Big]\Bigg\}.
\end{align}
Using
$\sin a \sin b = [\cos(a-b)-\cos(a+b)]/2$
and again taking $\ket{\psi}=\ket{x}$, we obtain
\begin{align} \nonumber
    \big[\mathrm{Im}(K_d(\omega_x,t))\big]^2
    =
    \frac{1}{d^2}\Big[&
    (d-1)^2 \frac{1-\cos(2\omega_x t)}{2}
    +
    \frac{1-\cos(2\omega_x' t)}{2} \\
    &-
    (d-1)\left(
    \cos[(\omega_x'-\omega_x)t]
    -
    \cos[(\omega_x'+\omega_x)t]
    \right)
    \Big].
\end{align}
Taking the Gaussian average with $\mu=0$ gives
\begin{align}\nonumber
     \overline{[\mathrm{Im}(K_d(\omega_x,t))]^2}
     =
     \frac{1}{d^2}\Big[&
     (d-1)^2 \frac{1-e^{-2\sigma^2\omega_x^2}}{2}
     +
     \frac{1-e^{-2\sigma^2\omega_x'^2}}{2} \\
     &-
     (d-1)\left(
     e^{-\frac{\sigma^2}{2}(\omega_x'-\omega_x)^2}
     -
     e^{-\frac{\sigma^2}{2}(\omega_x'+\omega_x)^2}
     \right)
     \Big].
\end{align}
Since $\ket{\psi}=\ket{x} \Longrightarrow \overline{[\mathrm{Im}(h_d(E,x|t))]} = \overline{[\mathrm{Im}(K_d(\omega_x,t))]^2}$, considering the limit $|\omega_x| \gg 1$, we obtain
\begin{align}
   \mathrm{Var}_m[\mathrm{Im}(h_d(E,x))]
   =
   \frac{(d-1)^2+1}{2d^2},
\end{align}
recovering the same result obtained for the real part.
The qubit implementation has no imaginary part, see eqn.~(\ref{H_echo_qubit}), so the result holds for $d>2$.

\section*{Author Declarations}

This research received no specific grant from any funding agency in the public, commercial, or not-for-profit sectors.

The authors used large language models, Claude (Anthropic) and ChatGPT (\mbox{OpenAI}), to improve the grammar and clarity of this manuscript. The authors prompted the AI to review flow and conciseness, but they verified and edited all content, interpretations, and conclusions and take full responsibility for the integrity of the work.

\subsection*{Conflict of Interest Statement}
The authors have no conflicts to disclose.

\subsection*{Author Contributions}

J.C.S.R. contributed to the conceptualization of the $d$-dimensional ancilla qudit formulation of the Rodeo algorithm, development of the methodology, and formal analysis. He was responsible for writing the original draft of the manuscript, implementing the numerical simulations, preparing the figures, and performing data curation and visualization. R.A.D. contributed to the conceptualization of the homogeneous input state and associated dataset, as well as to the development of data processing routines and visualization.
\section*{Data Availability Statement}
The data that support the findings of
this study are openly available in reference number~\cite{data_zenodo}.

\section*{References}
\bibliographystyle{unsrt}  
\bibliography{quditRodeo}

\end{document}